%% file: MAIN2.tex
\documentclass[10pt]{article}
\usepackage{authblk}
\usepackage{xspace}
\usepackage{enumerate}
\usepackage{times,epsfig,makeidx,amsfonts,amsmath}
\usepackage{natbib}
\newcommand{\Lcal}{\mathcal{L}}
\newcommand{\Var}{\mathbb{V}}

\newtheorem{theorem}{Theorem}

\newtheorem{lemma}[theorem]{Lemma}

\newcommand{\SegCorr}{\mbox{SegCorr}\xspace}

\newcommand{\EXP}{{Y}}

\newcommand{\ECNV}{ \EXP }
\newcommand{\TECNV}{ \EXP^{\top} }
\newcommand{\nECNV}{ \EXP }
\newcommand{\prob}[1]{\mathbb{P}\left(#1 \right)}
\newcommand{\gp}[1]{\left(#1\right)}

\newcommand{\SpaceBeforeSection}{\vspace{-.02\textheight}}
\date{}

\begin{document}

\title{\SegCorr: a statistical procedure for the detection of genomic regions of correlated expression.}
\author[1,2,3,4]{Eleni Ioanna Delatola  }
\author[1,2]{Emilie Lebarbier }
\author[1,2,5]{Tristan Mary-Huard \\}
\author[3,4]{Fran\c{c}ois Radvanyi}
\author[1,2]{St\'ephane Robin}
\author[3,4]{Jennifer Wong  }
\affil[1] {AgroParisTech UMR518, Paris 5e, France}
\affil[2] {INRA UMR518, Paris 5e, France}
\affil[3] {Institut Curie, Centre de Recherche, Paris, F-75248 France}
\affil[4] {CNRS UMR144, Equipe Oncologie Mol\'eculaire, Paris, F-75248 France}
\affil[5] {INRA, UMR 0320 / UMR 8120 G\'en\'etique V\'eg\'etale et \'Evolution       Le Moulon, F-91190 Gif-sur-Yvette, France}



\maketitle

\begin{abstract}
\noindent {\bf Motivation:} Detecting local correlations in expression between neighbor genes along the genome has proved to be an effective strategy to identify possible causes of transcriptional deregulation in cancer. It has been successfully used to illustrate the role of mechanisms such as copy number variation (CNV) or epigenetic alterations as factors that may significantly alter expression in large chromosomic regions (gene silencing or gene activation).\\
{\bf Results:} The identification of correlated regions requires segmenting the gene expression correlation matrix into regions of homogeneously correlated genes and assessing whether the observed local correlation is significantly higher than the background chromosomal correlation. A unified statistical framework is proposed to achieve these two tasks, where optimal segmentation is efficiently performed using dynamic programming algorithm, and detection of highly correlated regions is then achieved using an exact test procedure.
We also propose a simple and efficient procedure to correct the expression signal for mechanisms already known to impact expression correlation. The performance and robustness of the proposed procedure, called \SegCorr, are evaluated on simulated data. The procedure is illustrated on cancer data, where the signal is corrected for correlations possibly caused by copy number variation. The correction permitted the detection of regions with high correlations linked to DNA methylation.\\
%
{\bf Availability and implementation:} R package \SegCorr is available on the CRAN.
{\bf Contact:} eldelatola@yahoo.gr
\end{abstract}

\SpaceBeforeSection
\section{Introduction} \label{sec:intro}
\input{Intro.tex}

\SpaceBeforeSection
\section{Correlation matrix segmentation} \label{sec:region}
\input{StatModel2.tex}
\input{RegionInference3.tex}

\SpaceBeforeSection
\section{Assessing correlation significance} \label{sec:regiontest}
\input{Test.tex}
\SpaceBeforeSection
\section{Simulation study} \label{sec:simul}
\input{SimulationStudy3.tex}

\SpaceBeforeSection
\section{Bladder cancer data} \label{sec:bladder}
In this section, we apply \SegCorr on the dataset described in Section \ref{sec:data_presentation}. It is now well known that copy number
variation (CNV) impacts gene expression \citep{sebat2004large}. Here our goal is to detect regions where the
correlation is not due to CNV. Therefore we correct the expression signal for CNV variation according to the strategy described in Sections
\ref{sec:statmodel} and \ref{sec:Preprocessing}. The effect of this correction is investigated in
Section \ref{study-correction-CNV}. Lastly, Section \ref{methylation}  illustrates the biological results obtained after correction for CNV.

\input{DataPres.tex}
\input{SRobustness.tex}
\input{PrePro.tex}
\input{CNVregion.tex}
\input{MethylatedRegion.tex}

\SpaceBeforeSection
\section{Discussion} \label{sec:discussion}
\input{Discussion.tex}
\vspace{-.2cm}
\subsection*{Acknowledgements}
This work has been supported by the INCa\_4382 research grant. The authors thank E. Chapeaublanc from Institut Curie for providing the data.

\appendix
\SpaceBeforeSection
\section*{Appendix}
\renewcommand{\thesubsection}{\Alph{subsection}}
\input{Appendix.tex}
\bibliography{MAIN2}

\end{document}

%% file: Intro.tex

In the last decade, the study of local co-expression of neighbor genes along the chromosome has become a question of major importance in cancer biology \citep{clark2007action}. The development of "Omics" technologies have permitted the identification of several mechanisms inducing local gene regulation, that may be due to a common transcription factor \citep{De2010} or common epigenetic marks \citep{stransky2006regional,Frigola06}. Copy number variation due to polymorphism or to genomic instability in cancer is also a possible cause for observing a correlation between neighbor genes \citep{Aldred05}, as their expressions are likely to be affected by the same copy number variation (CNV). It has further been observed that local regulations may occur in specific nuclear domains, as the nuclear region is an environment which may favor or not transcription \citep{Bickmore13}. \\

Investigating the impact of a specific source of regulation (TF, DNA methylation, histoine modifications, CNV) on the expression has now become a common practice for which statistical tools are readily available. On the other hand, only a few methods have been proposed to focus on the direct analysis of gene expression correlation along the chromosomes. The direct analysis of correlations may have different purposes:
\begin{enumerate}[($i$) ]
\item one can aim at detecting all potential chromosomal domains of co-expression, then investigating to which extend known causal mechanism are responsible for the observed co-expression patterns,
\item one can aim at detecting chromosomal domains of co-expression where correlations are not caused by already known sources of regulation, in order to identify new potential mechanisms impacting transcription.
\end{enumerate}
Addressing problems $(i)$ and $(ii)$ is crucial to fully understand transcriptional deregulation and/or to model gene regulation.

We first consider problem $(i)$ and provide a precise definition of our purpose: one aims at identifying correlated regions, i.e. blocks of neighbor genes, the expression of which display correlations across patients that are significantly higher than expected. Indeed, it has been observed that background correlation between adjacent genes along the genome does exist. This should not be confounded with the co-expression that can be locally observed due to the aforementioned mechanisms. Note that this definition does not include detection of regions with biased gene expression as studied in \cite{nilsson2008improved}, \cite{xiao2009improved} or \cite{seifert2014autoregressive}, that account for potential correlations between adjacent genes but do not aim at detecting highly correlated regions.

Several approaches have already been proposed to tackle problem $(i)$. Clustering methods accounting for the chromosomal organization of the genes have been proposed in CluGene \citep{dottorini2013clugene} and DIGMAP \citep{yi2005coupled}. Sliding windows procedures have also been considered, such as G-NEST \citep{lemay2012g}, REEF \citep{coppe2006reef} or TCM \citep{reyal2005visualizing}. The principle is then to compute correlation scores for genes falling within the window, then to detect local peaks of high correlation scores. While these procedures have been successfully applied to cancer data, all tackle the detection of correlated region using heuristics. As such, they suffer from classical limitations associated with these techniques, including local optimum (for clustering algorithms) or detection instability according to the choice of the window size (for sliding windows).

It is now well known that the problem of finding regions in a spatially ordered signal can be cast as a segmentation problem, for which standard statistical models exist, along with efficient algorithms to find the globally optimal solution. According to our definition, the detection of correlated regions boils down to the block-diagonal segmentation of the correlation matrix between gene expressions. Such an approach has been proposed in image processing \citep{lu2013correlation}, finance \citep{lavielle2006detection} and bioinformatics for CNV analysis \citep{zhang2010cmds}, but to the best of our knowledge it has never been considered for the detection of correlated expression regions.\\


While problem $(i)$ can be addressed on the basis of only expression data, problem $(ii)$ requires the additional measurement of the signal one needs to account for. For example, consider that one seeks for locally expressed co-regulation events that are not due to copy number variations to include copy number changes due to polymorphism or genomic alteration observed in cancer. The strategy we adopt here consists in first correcting the expression data for potential CNV regulation, then in applying the procedure described to solve problem $(i)$ on the corrected signal. The corrected signal is obtained by regressing the initial expression signal on the CNV signal. Although quite simple, the strategy turns out to be efficient in practice. An alternative strategy would be to jointly model both the expression and the signals to correct for, and then propose within this framework a correction. Such a strategy would necessitate to adapt the modeling to the specific combination of signals one has at hand. In comparison, the regression procedure proposed here can be applied to any kind and any number of signals one needs to correct for. \\

The outline of the present article is the following. In Section \ref{sec:region} we propose a parametric statistical framework for the problem of correlated region identification. Finding regions of co-regulated genes can then be achieved by maximum likelihood inference (to find the boundaries of each region along with their correlation levels). An exact test procedure to assess the significance of the correlation with respect to background correlation is proposed in Section \ref{sec:regiontest}. We introduce a simple procedure to correct expression data beforehand for some known (and quantified) source of correlation. Because the background correlation level is a priori unknown, an estimator of this quantity is also proposed.
The performance of the \SegCorr procedure is illustrated in Section \ref{sec:simul} on simulated data, along with a comparison with the TCM algorithm proposed in \citet{reyal2005visualizing}. Finally, a case study on cancer data is presented in Section \ref{sec:bladder}.

%% file: StatModel2.tex
\subsection{Statistical model} \label{sec:statmodel}
We consider the following expression matrix:
$$
\nECNV =
\begin{bmatrix}
\nECNV_{11} & \cdots & \nECNV_{1p} \\
\nECNV_{21} & \cdots & \nECNV_{2p} \\
\vdots & \ddots & \vdots \\
\nECNV_{n1} & \cdots & \nECNV_{np}
\end{bmatrix}
$$ 
where $\nECNV_{ij}$ stands for the expression of gene $j$ ($j =
1,\ldots,p$) observed in patient $i$ ($i = 1,\ldots,n$). The $i$-th
row of this matrix is denoted $\nECNV_i$ and corresponds to the
expression vector of all genes in patient $i$.  In order to detect
regions of correlated expression, we consider the following
statistical model. Profiles $\{\nECNV_i\}_{1 \leq i \leq n}$ are
supposed to be i.i.d, normalized (centered and standardized), following a Gaussian distribution with
block-diagonal correlation matrix $G$:
\begin{eqnarray}\label{eq:CorrelationModel}
G &=&
\begin{bmatrix}
  \Sigma_1 &  &  &  &  \\
  & \ddots &  & \bf{0} &  \\
  &  & \Sigma_k &  &  \\
  &  \bf{0} &  & \ddots &  \\
  &  &  &  & \Sigma_K \\
\end{bmatrix} \nonumber\\
\text{ with }
\Sigma_k &=&
\begin{bmatrix}
  1 & \cdots & \rho_{k} \\
  \vdots & \ddots & \vdots \\
  \rho_{k} & \cdots & 1
\end{bmatrix}.
\end{eqnarray}
The model states that genes are spread into $K$ contiguous regions,
with respective lengths $p_k$ ($k = 1, \ldots, K$, $\sum_{1 \leq k
  \leq K} p_k = p$). Genes belonging to different regions are supposed
to be independent, whereas genes belonging to a same region are
supposed to share the same pairwise correlation coefficient
$\rho_k$. This amounts to assume that some specific effect
  (e.g. methylation) affects the expression of all genes belonging to
  the region. More specifically, let $U_k$ denote the vector of the
    region effect (accross patients). For all genes $j$ from region
    $k$, the model can be written as $Y_{ij} = \mu_{ij} + U_{ik} +
    E_{ij}$. The error terms $E_{ij}$ are all independent and
    independent from $U_{ik}$ such that $\Var(U_{ik})/\Var(Y_{ij}) =
    \rho_k$.

\subsection{Accounting for known sources of regulation}
As mentioned in the Introduction, a second task $(ii)$ can be to
detect correlated regions which are not due to an already known
mechanism. While in the model of the previous
  section, $Y_{ij}$ denoted the expression of gene $j$ for patient
  $i$. In this case, $Y_{ij}$ refers to the gene expression corrected
  for some underlying mechanism. This correction step can be done
using the following regression model~:
\begin{equation}
\nECNV_{ij} = \beta_{0} + \beta_{1} x_{ij} + \epsilon_{ij},
\label{eq:regEXP1}
\end{equation}
where $x_{ij}$ stands for the covariate observed in patient $i$ for
gene $j$. For instance, in the illustration of Section
\ref{study-correction-CNV}, $x_{ij}$ is the copy number associated to
patient $i$ at location of gene $j$. The corrected signal is then
$\widetilde{\nECNV}_{ij}=\nECNV_{ij}- \widehat{\beta}_{0} -
\widehat{\beta}_{1} x_{ij}$. Note that it suffices to assume
  that $(\epsilon_{ij})$ are independent among patients (but not among
  genes) to get the standard linear regression estimates (see \cite
  {And58}, Chapter 8).

Several articles have investigated the
relationship between gene expression and mechanisms such as CNV or
methylation, and proposed sophisticated models for this relationship
\citep{menezes2009integrated,van2010random,leday2013modeling}. Alternatively, gene
  expression can be predicted using one of these models. Then, these
  predictions can replace the explanatory variable $x$ in Equation
  \eqref{eq:regEXP1}. The residuals of the resulting model can be
  considered as the corrected signal. 

%% file: RegionInference3.tex
\subsection{Inference of correlated regions} \label{sec:regioninference}

Parameter inference in Model \eqref{eq:CorrelationModel} amounts to estimate the number of regions $K$, the
region boundaries $0 = \tau_0 < \tau_1 < \dots < \tau_{K+1}
= p$, and the correlation parameters $\rho_1,...,\rho_K$ within each of these regions. Here, we
consider a maximum penalized likelihood approach. First, we show that for a given $K$ the optimal region boundaries and correlation coefficients can be efficiently obtained using dynamic programming. The number of regions can then be selected using a penalized likelihood criterion. \\
For a fixed $K$, the estimation problem can be formulated as follows:
\begin{eqnarray} \label{eq:InferenceKfixed}
\arg\max_{\tau_1 < \dots < \tau_K} \max_{\rho_1, \ldots,\rho_K}
\Lcal
\end{eqnarray}
where the log-likelihood $\Lcal$ is expressed as
\begin{eqnarray*}
-2 \Lcal & = & n\log|G| + \mbox{tr}\left[ \nECNV G^{-1}
(\nECNV)^\top \right]. .
\end{eqnarray*}
Here, thanks to the block diagonal structure of the correlation
matrix in Model \eqref{eq:CorrelationModel}, the log-likelihood can
be rewritten as 
\begin{eqnarray*}
-2 \Lcal & = & \sum_{k=1}^K \left\{ n\log|\Sigma_k| +
\mbox{tr}\left[ \nECNV^{(k)} \Sigma_k^{-1} (\nECNV^{(k)})^\top
\right] \right\}
\end{eqnarray*}
where $\nECNV^{(k)}$ stands for the set of expression from $\nECNV$
corresponding to genes included in the $k$-th region. 
Moreover defining $\Lcal_k$, the log-likelihood in region $k$ containing genes
from $\tau_{k-1}+1$ to $\tau_k$ as  
\begin{eqnarray} \label{eq:Lk}
- 2 \Lcal_k &=&  -2\Lcal(\tau_{k-1}+1,\tau_k)\\
&=& n \log|\Sigma_k| + \mbox{tr}\left[ \nECNV^{(k)} \Sigma_k^{-1} (\nECNV^{(k)})^\top \right] \ ,
\end{eqnarray}
the optimization problem \eqref{eq:InferenceKfixed} boils down to
\begin{eqnarray} \label{eq:InferenceKfixed1}
\arg\max_{\tau_1 < \dots < \tau_K} \sum_{k=1}^K \max_{\rho_k}
\Lcal_k.
\end{eqnarray}

\paragraph{Inference when $K$ is fixed} 
We first show that for a given region $k$ with known boundaries, explicit expressions can be obtained for both the ML estimator $\widehat{\rho}_k$ and the likelihood $\Lcal_k$ at the optimum:
\begin{lemma} \label{lem:rho}
The maximum of $\Lcal_k$ with respect to $\rho_k$ is reached
for
\begin{equation} \label{eq:rhok_hat}
\widehat{\rho}_{k} =
\frac{\sum_{j}^{p_k}{\sum_{k}^{p_k}{\widehat{G}_{jk}}}-p_k}{p_k^2 - p_k}
\end{equation}
where $\widehat{G}_{jk} :=  n^{-1} \sum_{i=1}^{n}\nECNV_{ij}\nECNV_{ik}$.
Furthermore, the maximal value of $\Lcal_k$ is given by
\begin{eqnarray}  \label{eq:contrast}
-2 \widehat{\Lcal}_k &=&
n\left[p_k + (p_k-1)\log{\left(-\frac{\sum_{j}^{p_k}{\sum_{k}^{p_k}{\widehat{G}_{jk}}}-p_k^2}{p_k^2 - p_k} \right)} \right.\nonumber\\
&& \left. +
\log{\left(\frac{\sum_{j}^{p_k}{\sum_{k}^{p_k}{\widehat{G}_{jk}}}}{p_k}
\right)}\right].
\end{eqnarray}
\end{lemma}
The proof is given in Appendix \ref{annex: lemma}. The expression of Problem \eqref{eq:InferenceKfixed1} is now 
\begin{equation} \label{eq:contrastK}
\arg\max_{\tau_1 < \dots < \tau_K} \sum_{k=1}^K \widehat{\Lcal}_k \ 
\end{equation}
which is additive with respect to the $\widehat{\Lcal}_k$ terms that can be straightforwardly computed thanks to Lemma \ref{lem:rho}. Consequently, optimization can be performed via Dynamic Programming (DP, \cite{Lavielle20051501}, \cite{PRL05}). The optimal boundaries, and correlation estimators can be obtained at computational cost $\mathcal{O}(Kp^2)$. \\
Lasso-type approaches have been proposed to tackle segmentation problems in a faster way (see e.g. \cite{tibshirani2008spatial}). First, note that such methods rely on a relaxation of the original problem, so that the result may be different from the exact solution of problem \eqref{eq:contrast}. Furthermore, as for matrix segmentation, such approaches have been proposed (\cite{bien2011sparse,levina2008sparse}), which do not allow to capture the longitudinal structure (i.e. blocks of neighbor genes).

\paragraph{Model selection.}\label{modelselection}
To choose the number of regions, we adopt the model selection
strategy proposed in \citet{Lavielle20051501}. For each $1\leq K
\leq K_{\max}$, we define the maximal log-likelihood for $K$ regions as
$$
L_K =  \max_{\tau_1 < \dots < \tau_K} \sum_{k=1}^K
\widehat{\Lcal}(\tau_{k-1}+1, \tau_k) \ .
$$
Furthermore, the normalized log-likelihood is defined as
$$
\widetilde{L}_{K} = \frac{L_{K_{\max}} - L_{K}}{L_{K_{\max}} -
L_{1}}(\widetilde{K}_{\max} - \widetilde{K}_{1}) + 1,
$$
where $\widetilde{K}_{j} = 5\times j + 2\times j \log{(p/j)}$ is the
penalty function. \citet{Lavielle20051501} suggests to estimate the
number of regions $\widehat{K}$ as the value of $K$ such that
$\widetilde{L}_{K}$ displays the largest slope change. Namely, we
take
\begin{displaymath}
\widehat{K} = \arg\min_K \left\{ (\widetilde{L}_{K} -
\widetilde{L}_{K+1}) - (\widetilde{L}_{K+1} - \widetilde{L}_{K+2}) >
S \right\},
\end{displaymath}
where  the value of threshold $S$ is predefined. Throughout the paper, $S=0.7$ as suggested in \cite{Lavielle20051501}. The robustness of the results with respect to other values for threshold $S$ is investigated in Section
\ref{sec:simul}. This global approach (dynamic programming and model selection) has been applied with success for CNV detection (see \cite{PRL05} and \cite{LJK05} for a comparative study.)

%% file: Test.tex
It has been observed \citep{cohen2000computational,spellman2002evidence,reyal2005visualizing,stransky2006regional} that background
correlations may exist between adjacent genes along the genome, i.e.
one expects the correlation level in any region to be positive. As a
consequence, one has to check whether a given region exhibits a
correlation level that is significantly higher than the background
correlation level $\rho_0$, that is observed by default.


\paragraph{Test procedure.}
Once the correlation matrix segmentation is performed, it is possible to identify regions with high correlation levels by testing $H_{0}: \rho_{k}=\rho_{0}$ vs $H_{1}: \rho_{k}>\rho_{0}$. This can be done using the following test statistic for region $k$:
\begin{eqnarray*}
T_k=\frac{1}{n}\sum_{i}^{n}  \gp{Y^{(k)}_{i\bullet} - Y^{(k)}_{\bullet\bullet}}^{2}
\end{eqnarray*}
where
\begin{eqnarray*}
Y^{(k)}_{i\bullet} = \sum_{j =\tau_{k-1}+1}^{\tau_{k}} Y_{ij}\ \text{ and } \ Y^{(k)}_{\bullet\bullet} = \sum_{i =1}^{n} Y^{(k)}_{i\bullet} \ .
\end{eqnarray*}
Assuming Model \eqref{eq:CorrelationModel} is true, test statistic $T_k$ has distribution
\begin{eqnarray*}
T_k \sim \lambda(p_k,\rho_k) \chi^{2}_{n-1} \ \text{ where } \ \lambda(p_k,\rho_k)=\frac{(1 + (p_{k}-1)\rho_{k})}{np_{k}} \ .
\end{eqnarray*}
Here $\chi^2_{n-1}$ stands for the chi-square distribution with $n-1$ degrees of freedom. The proof is given in Appendix \ref{annex: test}. We emphasize that this test is exact and does not rely on any resampling strategy. \\
Consequently, the $p$-value associated to region $k$ is given by
\begin{eqnarray*}
\prob{\lambda(p_k ,\rho_0)Z > T_k^{obs}}\ \ , \text{ where } \ \ Z\sim \chi^{2}_{n-1}.
\end{eqnarray*}

\begin{figure*}
\begin{center}
\begin{tabular}{ccc}
\includegraphics[width=0.30\textwidth, height=0.4\textwidth]{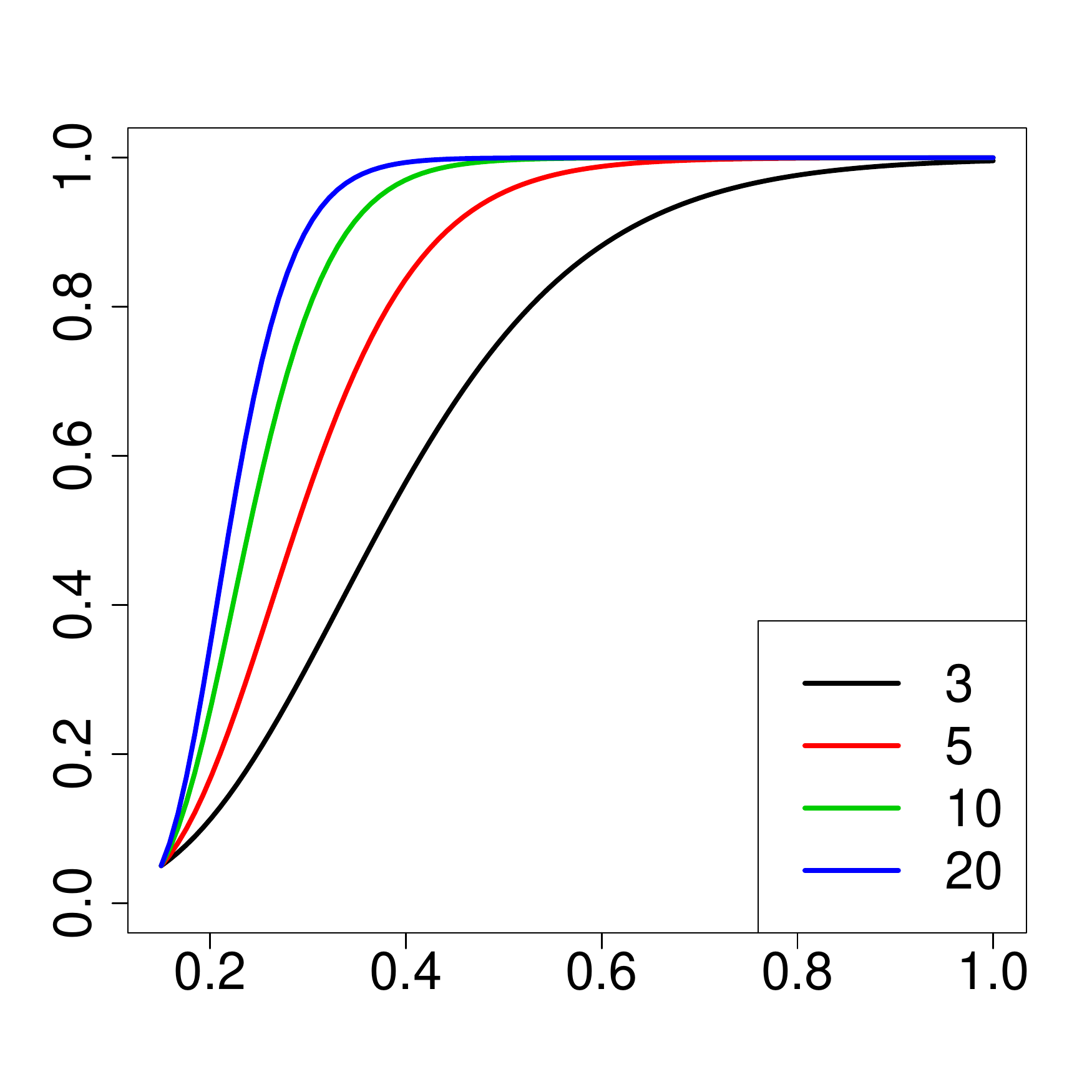}&
\includegraphics[width=0.30\textwidth, height=0.4\textwidth]{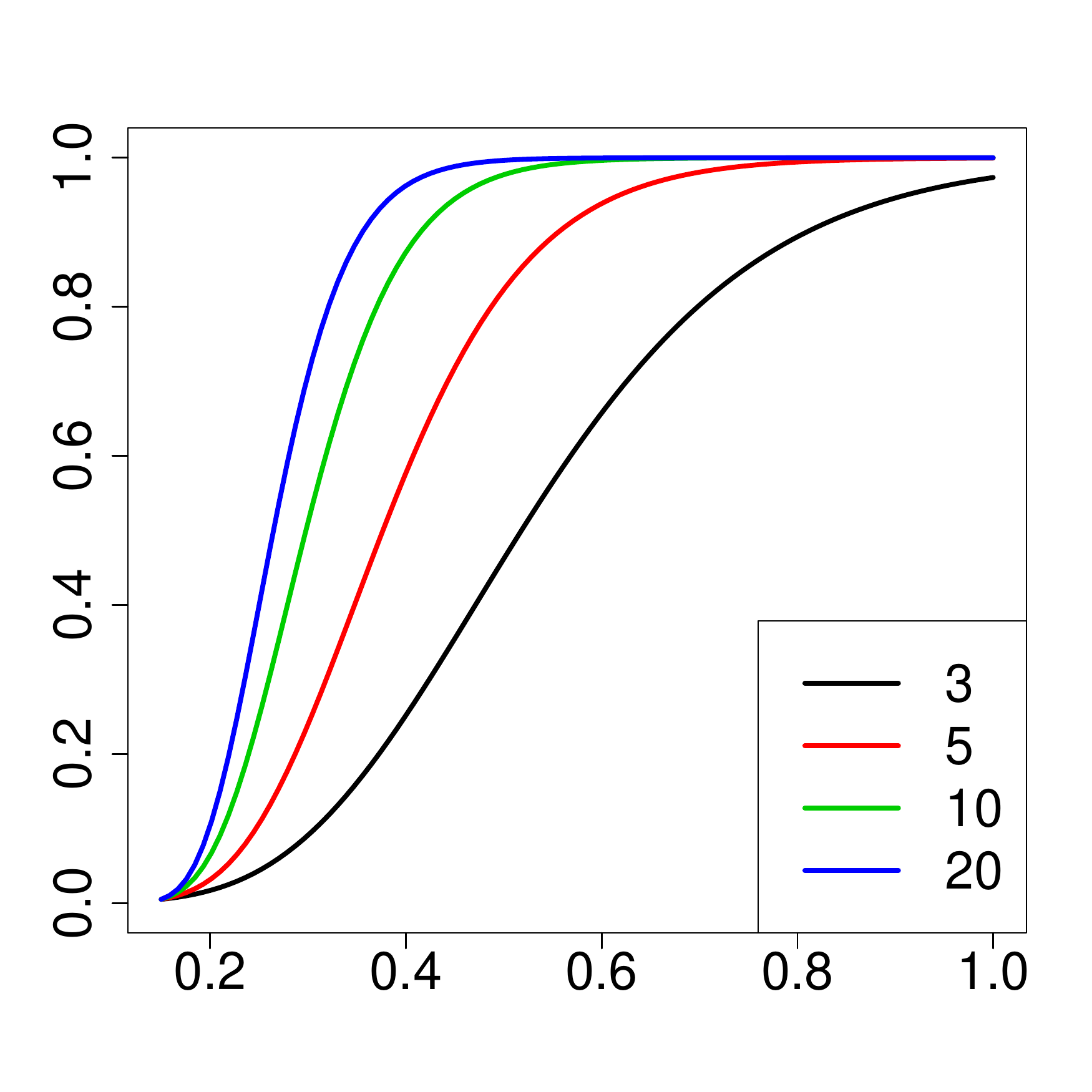}&
\includegraphics[width=0.3\textwidth, height=0.4\textwidth]{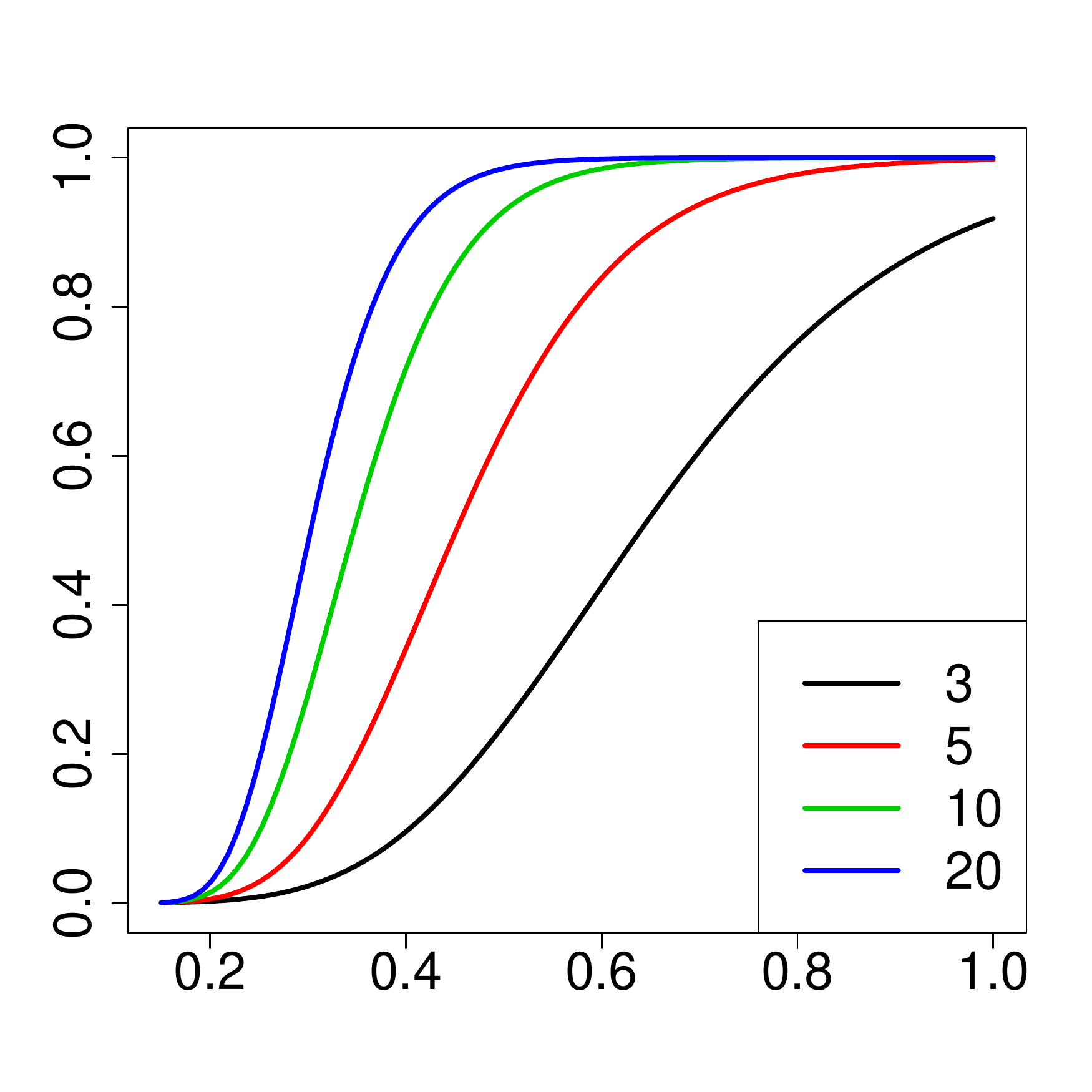}\\
\includegraphics[width=0.3\textwidth, height=0.4\textwidth]{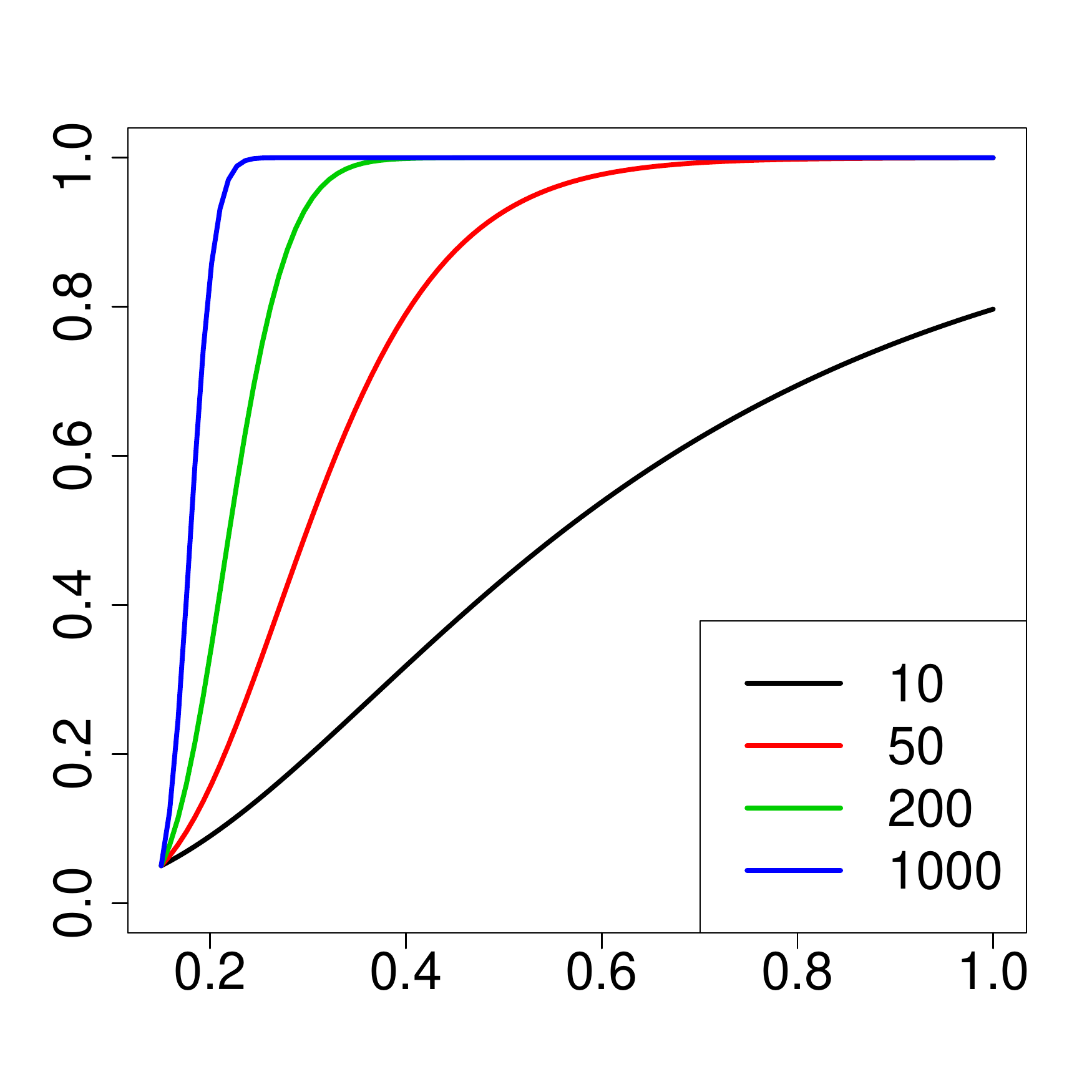}&
\includegraphics[width=0.3\textwidth, height=0.4\textwidth]{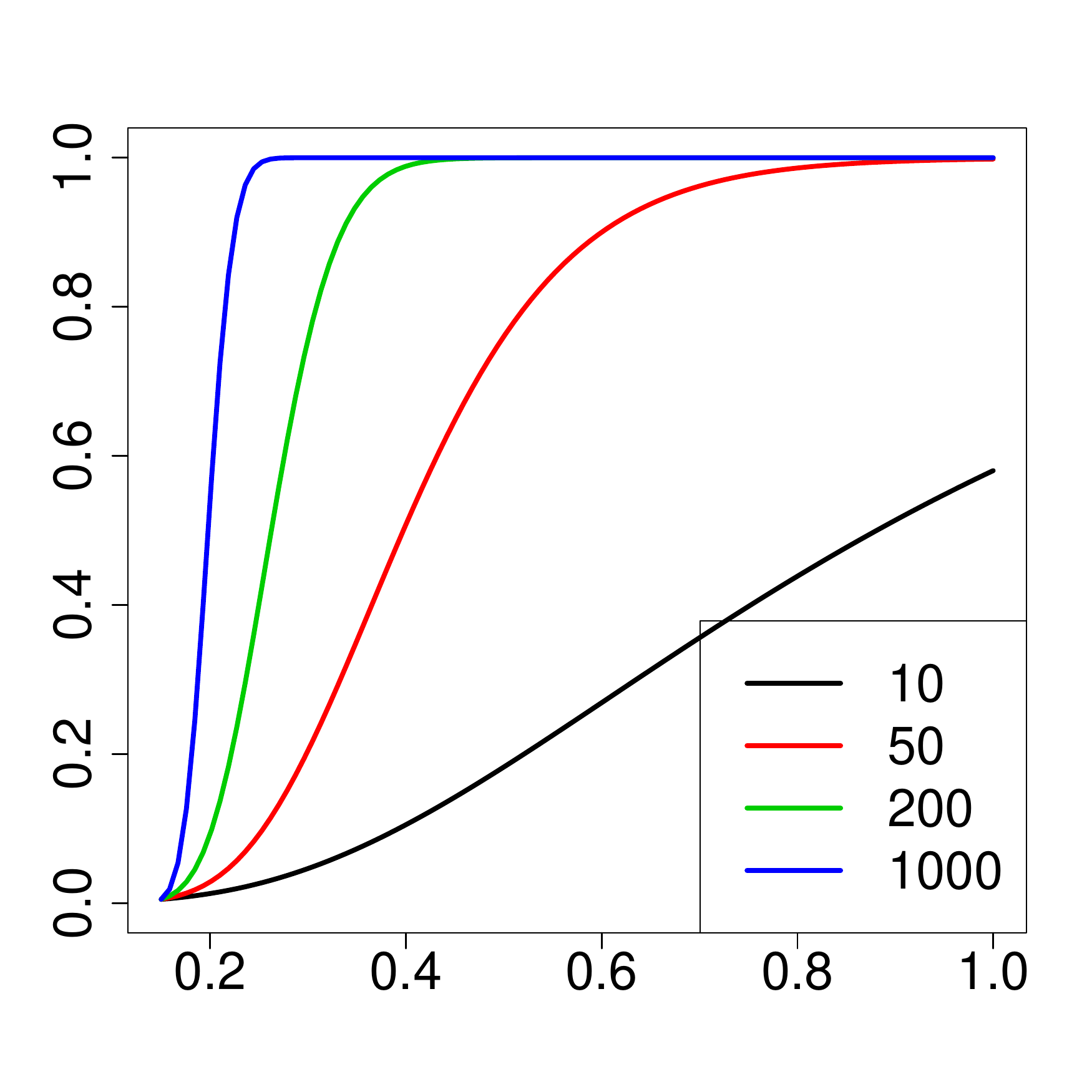}&
\includegraphics[width=0.3\textwidth, height=0.4\textwidth]{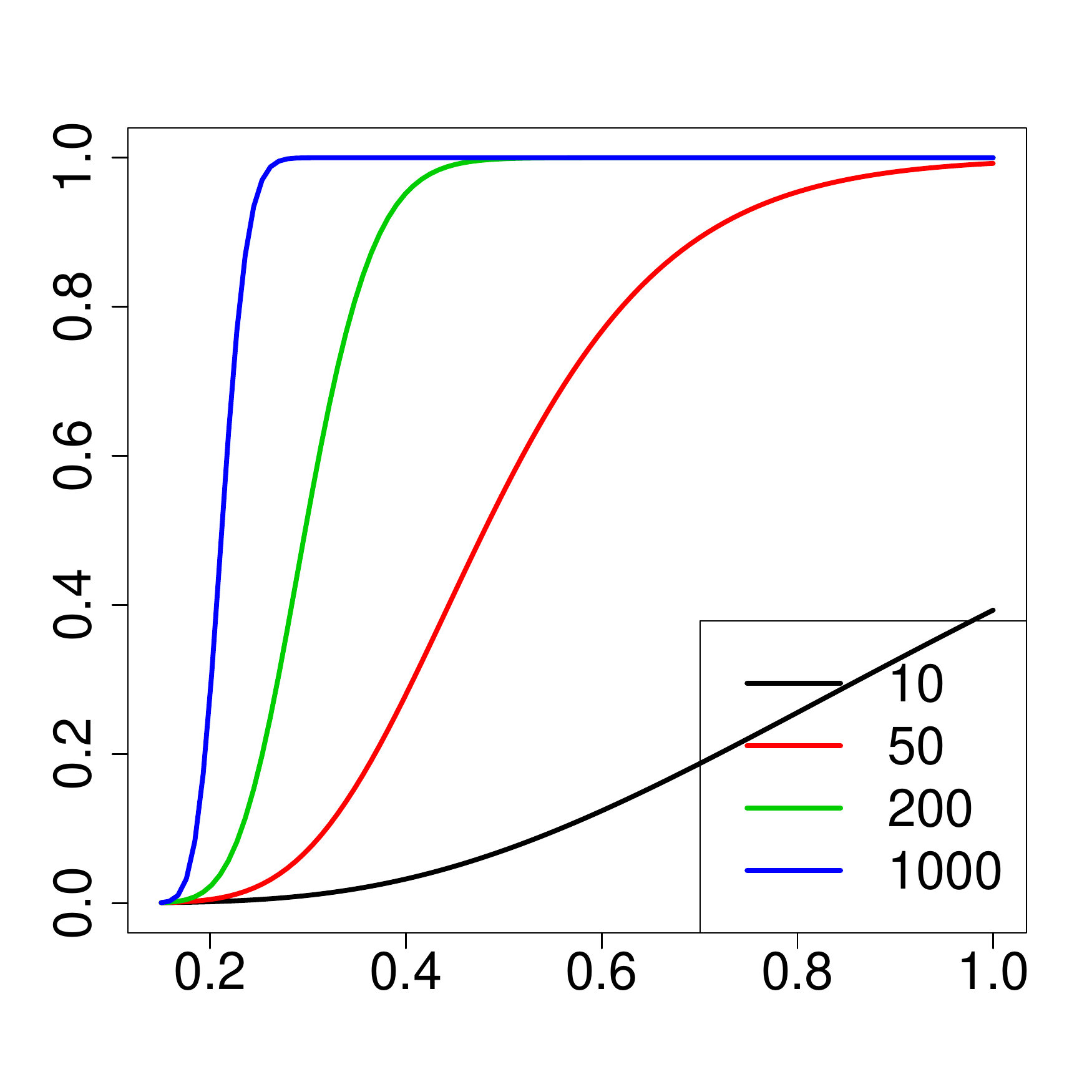}\\
\end{tabular}
\end{center}
\caption{{Top:} Power curves as a function of $\rho$, for a fixed cohort size $n=58$ and varying region width $p = 3, 5, 10, 20$. {Bottom:} Same graphs for a region of fixed width $p=5$ but varying cohort sizes $n=10, 50, 200, 1000$. In all graphs $\rho_0$ is fixed at 0.15. The nominal level $\alpha$ of the test is set to 5\% (left), 0.5\% (center), 0.05\% (right)\label{Figure:TheoreticalPower}}
\end{figure*}

\paragraph{Statistical power.} We now study the ability of the proposed test to detect a region with width $p$ where the correlation $\rho$ is higher than in the background. The probability to detect such a region depends on both $p$ and $\rho$, and writes
\begin{eqnarray*}
Po(n,p,\rho) &=& \Pr\{T > \lambda(p, \rho_0)q_{n-1,1-\alpha}\}\\
&=&  \Pr\left\{ Z > \frac{\lambda(p, \rho_0)}{\lambda(p, \rho)}q_{n-1,1-\alpha}\right\}
\end{eqnarray*}
where $Z\sim \chi^2_{n-1}$ and $q_{n-1,1-\alpha}$ is the $1-\alpha$ quantile for the  $\chi^2_{n-1}$ distribution. Figure \ref{Figure:TheoreticalPower} (Top) displays the evolution of power for different values of $p$ and $\rho$. Here $\rho_0$ and $n$ are fixed at 0.15 and 58, respectively, which correspond to the values observed in the reference dataset (see Section \ref{sec:bladder}). The nominal levels of $\alpha$ are 5\%, 0.5\%  and 0.05\%. These correspond to realistic thresholds, once multiple testing corrections such as Bonferroni or FDR are performed. One can observe that even for small values of $\rho$, the power is high whatever the nominal level as soon as the number of genes in the considered region is equal to or higher than 5. It also shows that the procedure will probably fail to find regions of size 3, if the correlation is not at least as high as 0.7 (to obtain a power of 0.8). On the same graph (Bottom), one observes that a sample of size 50 is sufficient to efficiently detect regions of size 5, as long as the
correlation is higher than 0.6. Larger samples will be required if one wants to efficiently detect regions with smaller correlation levels.

\paragraph{Background correlation estimation.}
The test procedure requires the knowledge of parameter $\rho_{0}$ that is unknown in practice. However, it can be estimated using
\begin{eqnarray}
\widehat{\rho_{0}}= |\underset{i>1}{\mbox{median}}(\mbox{corr}(\nECNV_{i-1},\nECNV_{i}))| \ . \label{eq: BackgroundCorrelationEstimator}
\end{eqnarray}
Under the assumption that most pairs of adjacent genes display a $\rho_{0}$ correlation, i.e. only a few number of regions with moderate sizes exhibit a high level of correlation, $\widehat{\rho_{0}}$ is a robust estimator of the background correlation. The behavior of estimator \eqref{eq: BackgroundCorrelationEstimator} is investigated in Section \ref{sec:simul}.

%% file: SimulationStudy3.tex
In this section, we first study the quality of the proposed estimator of
$\rho_0$. Then we study the ability of \SegCorr to detect correlated regions and compare its performance with this of TCM algorithm.
The robustness of the method with respect to the choice of
the model selection threshold $S$ will be investigated in Section
\ref{sec:SRobustness} on real data, since very little difference were
observed on the simulated data (results not shown).

\subsection{Simulation design}

\begin{figure}
  \includegraphics[width=0.45\textwidth, height=0.5\textwidth]{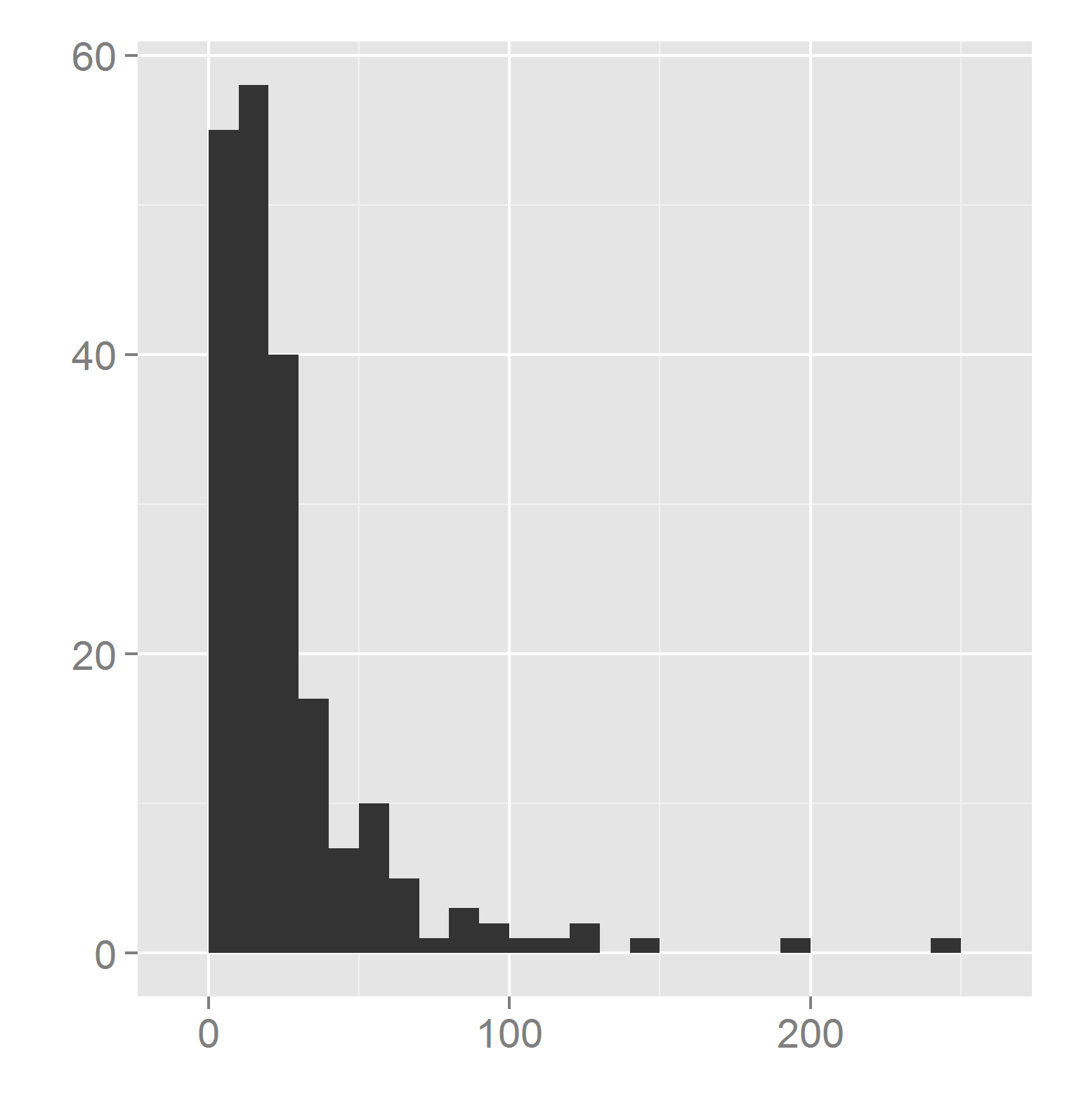}
  \includegraphics[width=0.45\textwidth, height=0.5\textwidth]{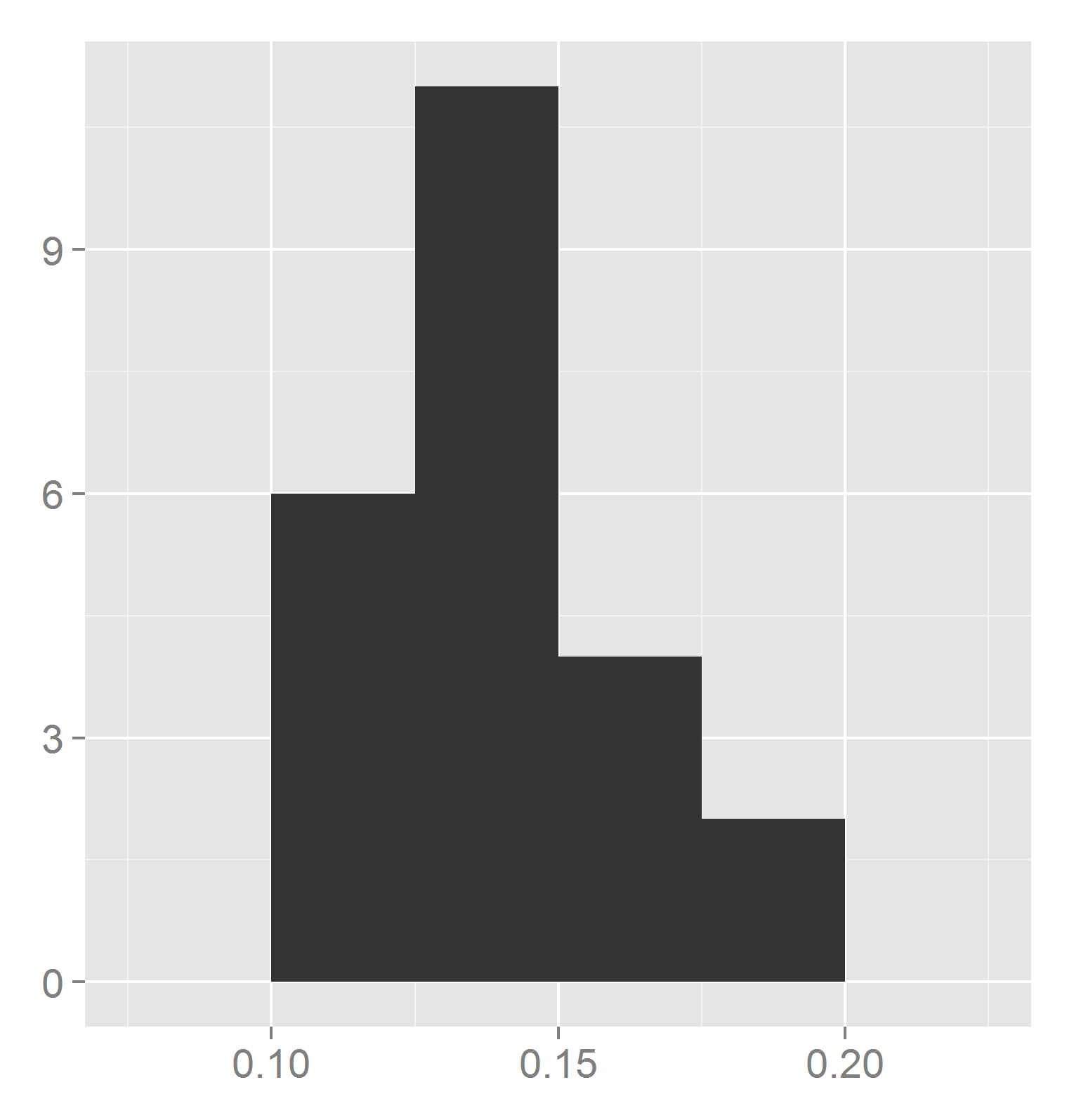}
  \caption{Left:~Length of $H_1$ regions in the reference dataset. Right:~Distribution of the
  backgroung correlation $\hat{\rho}_0$ obtained from the reference data according to the
  segmentation obtained in \citet{stransky2006regional}.}
  \label{fig:rho0}
\end{figure}

For each round of simulation, a sample of $n=58$
profiles with 22 chromosomes each is generated, with
a number of genes per chromosome identical to the one observed in the reference dataset (see description in Section \ref{sec:data_presentation}). Each chromosome is split into
$K_{chr}$ regions, according to the segmentation obtained in
\citet{stransky2006regional}. Each segmentation alternates between
$H_0$ regions, i.e. regions with nominal background correlation
$\rho_0$, and $H_1$ regions with nominal correlation $\rho_1$.
Figure~\ref{fig:rho0} depicts the length of the $H_1$ correlated
regions obtained in the \citet{stransky2006regional} study. As it can be seen, this
length varies substantially from a $H_1$ region to another. For
$\rho_1$, different values were considered (ranging from 0.3 to
0.9), all $H_1$ regions of all chromosomes sharing this same
$\rho_{1}$ coefficient. For $\rho_0$, two cases are considered~:
\begin{itemize}
\item Scenario 1 (Easy case):~ $\rho_0$ is the same for all $H_0$ regions of all
chromosomes (3 different values considered: 0.08, 0.18, 0.28).
\item Scenario 2 (Realistic case):~ $\rho_0$ is identical for all $H_0$ regions within a chromosome, but varies from a chromosome to another. The specific value chosen for a given chromosome is the one obtained for this same chromosome on the reference dataset.
The distribution of $\rho_{0}$ across chromosomes on the reference dataset is given in Figure~\ref{fig:rho0}.
\end{itemize}
Additionally, correlations between genes from different regions were observed on
the reference dataset. Thus, a similar extra block diagonal correlation pattern was generated in the simulations, with a level of
background correlation between blocks similar to the one within blocks.

Lastly, for each combination $(\rho_0,\rho_1)$
the simulation was replicated $20$ times.

\subsection{Quality of the $\rho_0$ estimator}
\begin{figure*}
\begin{center}
\includegraphics[width=.49\textwidth, height=0.5\textwidth]{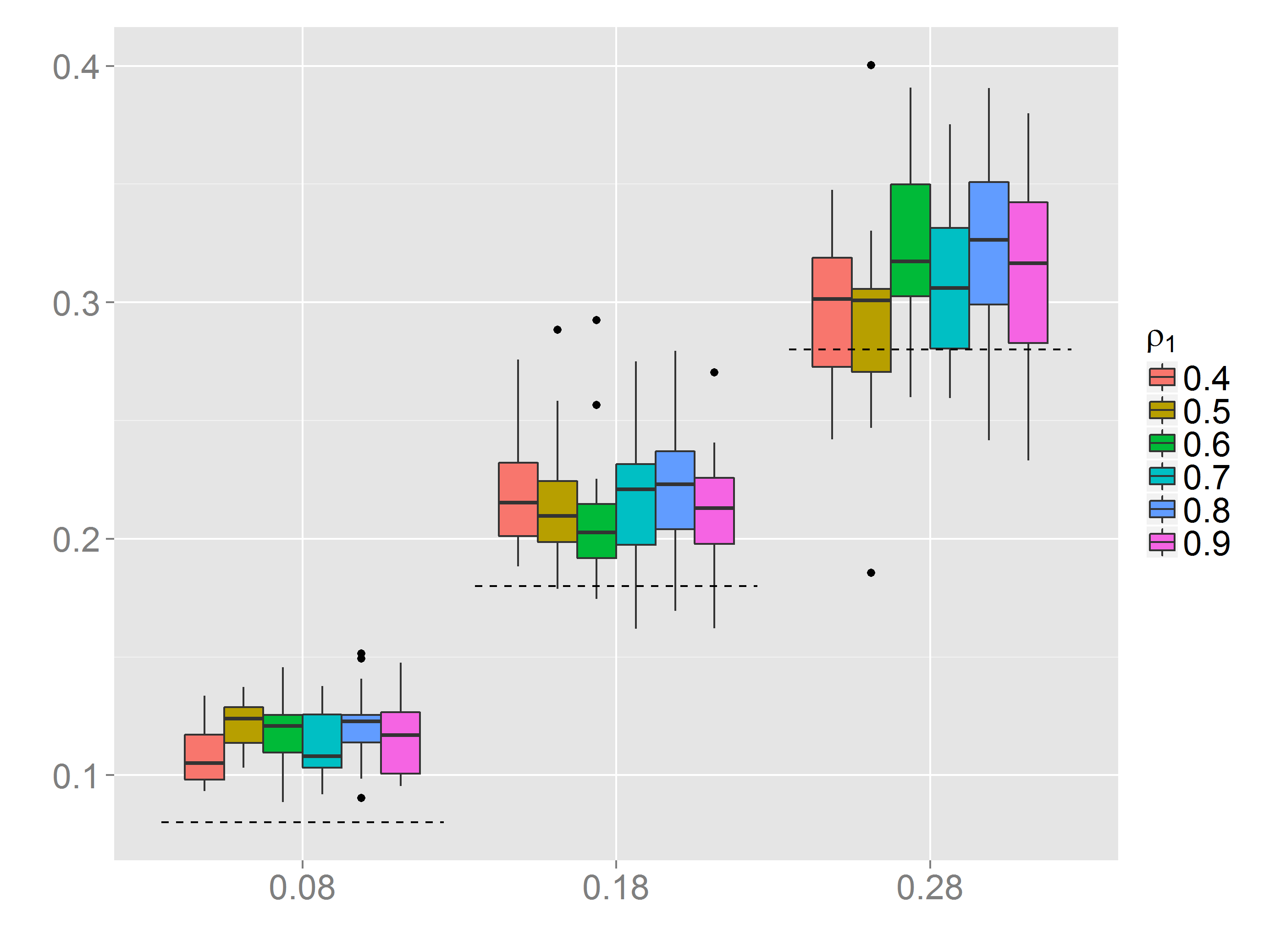}
\includegraphics[width=.49\textwidth, height=0.5\textwidth]{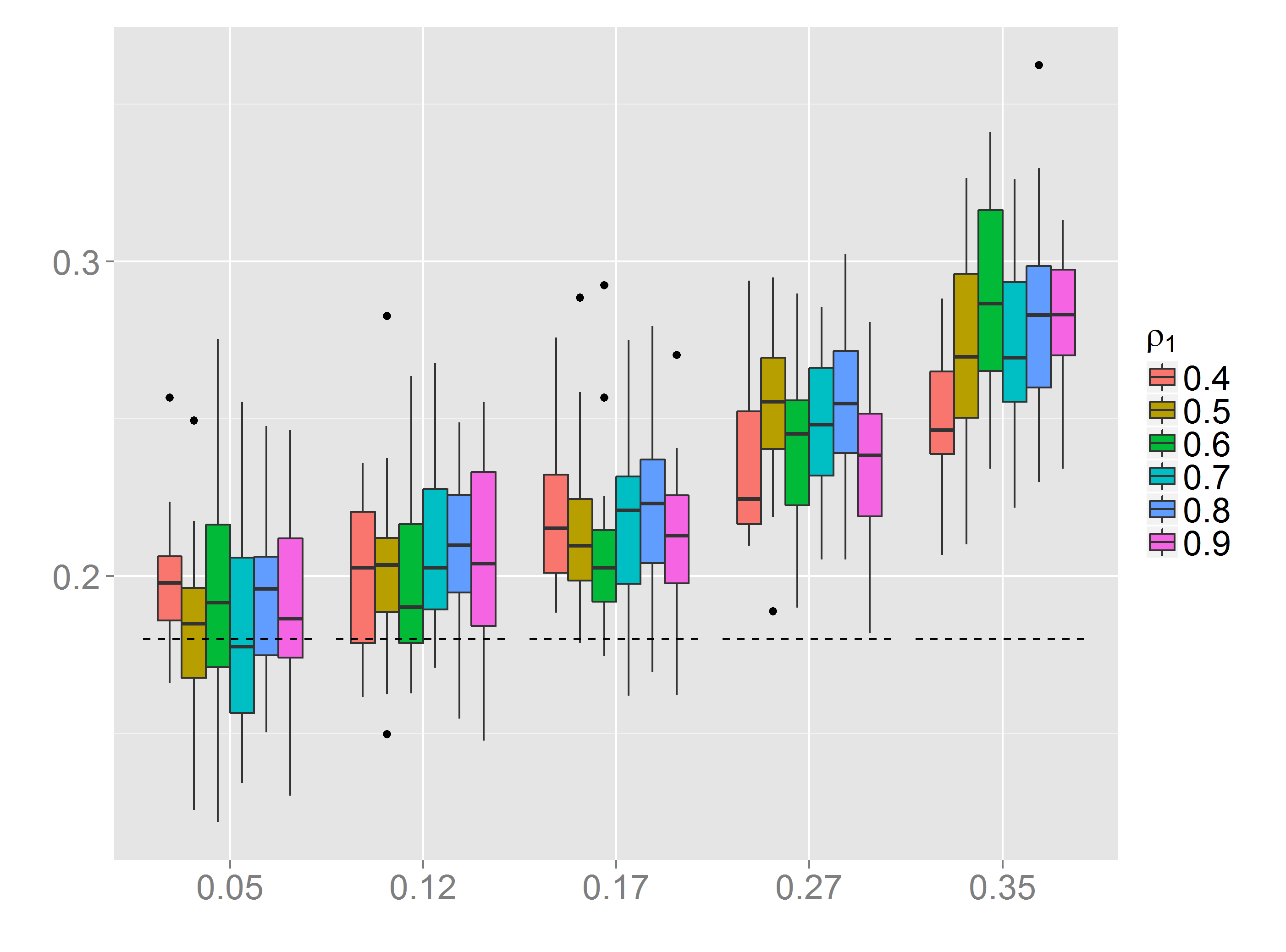}
\end{center}
\caption{Left: estimation of $\rho_0$ for chromosome 3 under
different levels of both $H_0$ and $H_1$ correlations
($\rho_0=0.08,0.18$ and $0.28$). Dashed lines indicate the true
$\rho_{0}$. Right: estimation of $\rho_0$ for $\rho_0=0.18$ and
different levels of $H_1$ correlations according to the fraction of
$H_1$ correlations (the results are showed for five typical chromosomes
only).} \label{fig:rho0hat}
\end{figure*}

For this study, we consider Scenario 1. Figure~\ref{fig:rho0hat}
illustrates the estimation accuracy of $\rho_{0}$ under
different levels of both $H_0$ and $H_1$ correlations on chromosome
3. Estimator \eqref{eq: BackgroundCorrelationEstimator} yields in
over-estimated values of the true background correlation level. One observes that the overestimation does not depend on
the correlation level in $H_1$ regions, thanks to the use of the
median. Still, as expected, it is linked to the proportion of pairs of
adjacent genes with $H_1$ correlations, as showed in
Figure~\ref{fig:rho0hat}. Importantly, while over-estimation of $\rho_0$ will result in a
decrease of power, it will not increase the false positive rate (FDR
or FWER).

\subsection{Performance evaluation}
To assess the performance of \SegCorr, the true positive
rate (TPR $=$ sensitivity), false positive rate (FPR $= 1-$
specificity) and area under the ROC curve~(AUC) were considered. These
criteria were first computed at the gene level. However,
as the goal is to identify correlated regions, a definition of TPR and FPR at the region level was adopted. We
considered the intersection between the true and the estimated
segmentations and computed the number of true/false positive/negative
regions. This amounts at classifying each gene into one of four
status (true/false $\times$ positive/negative) and then to merge
neighbor genes sharing a same status into regions. The status of a region is given by the
status of its genes. Consequently, criteria computed at
the region level are more stringent as they measure the precision of
region boundary estimation.

\begin{figure*}
	\begin{center}
  \includegraphics[width=0.45\textwidth, height=0.5\textwidth]{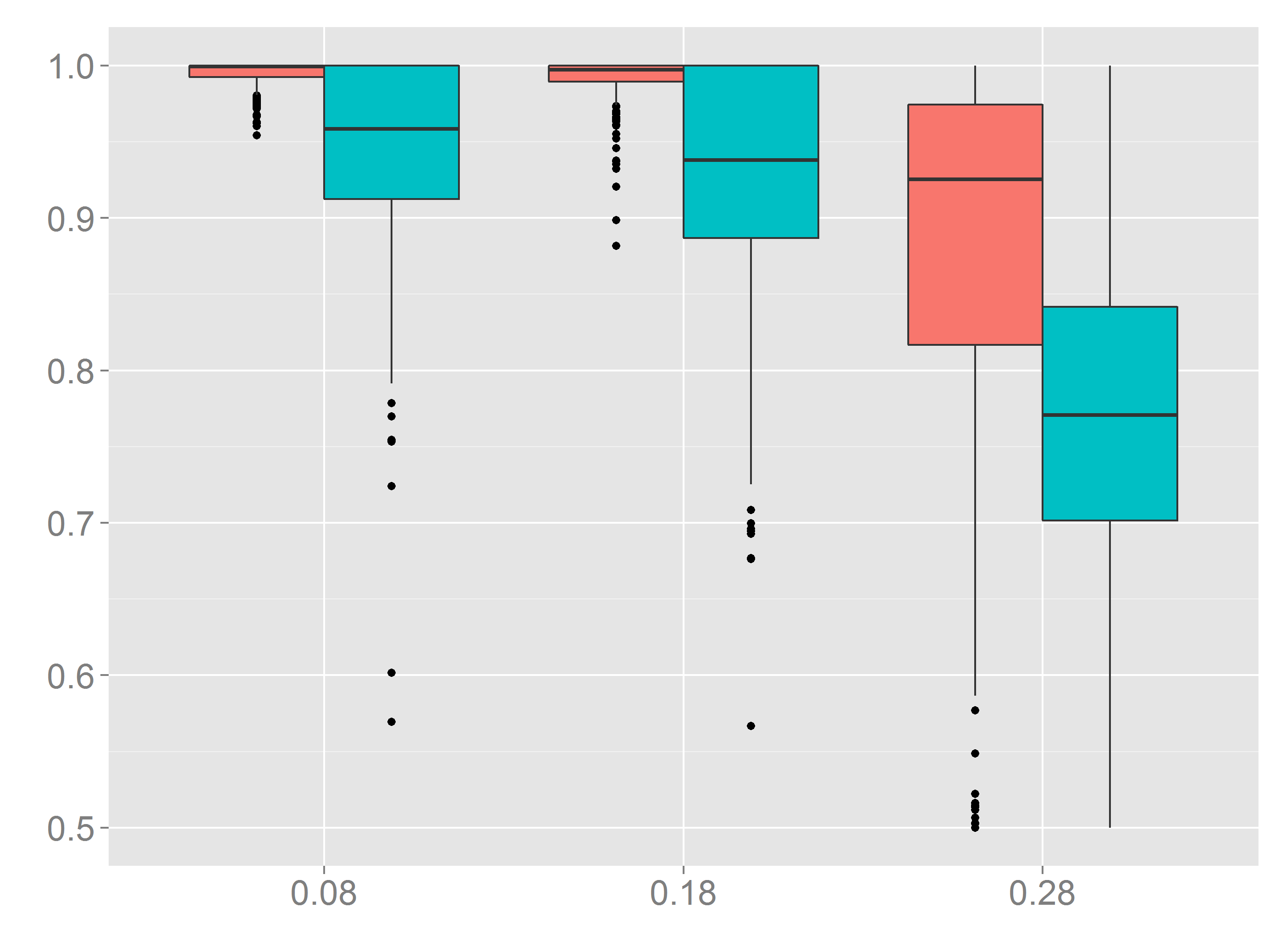}
  \includegraphics[width=0.45\textwidth, height=0.5\textwidth]{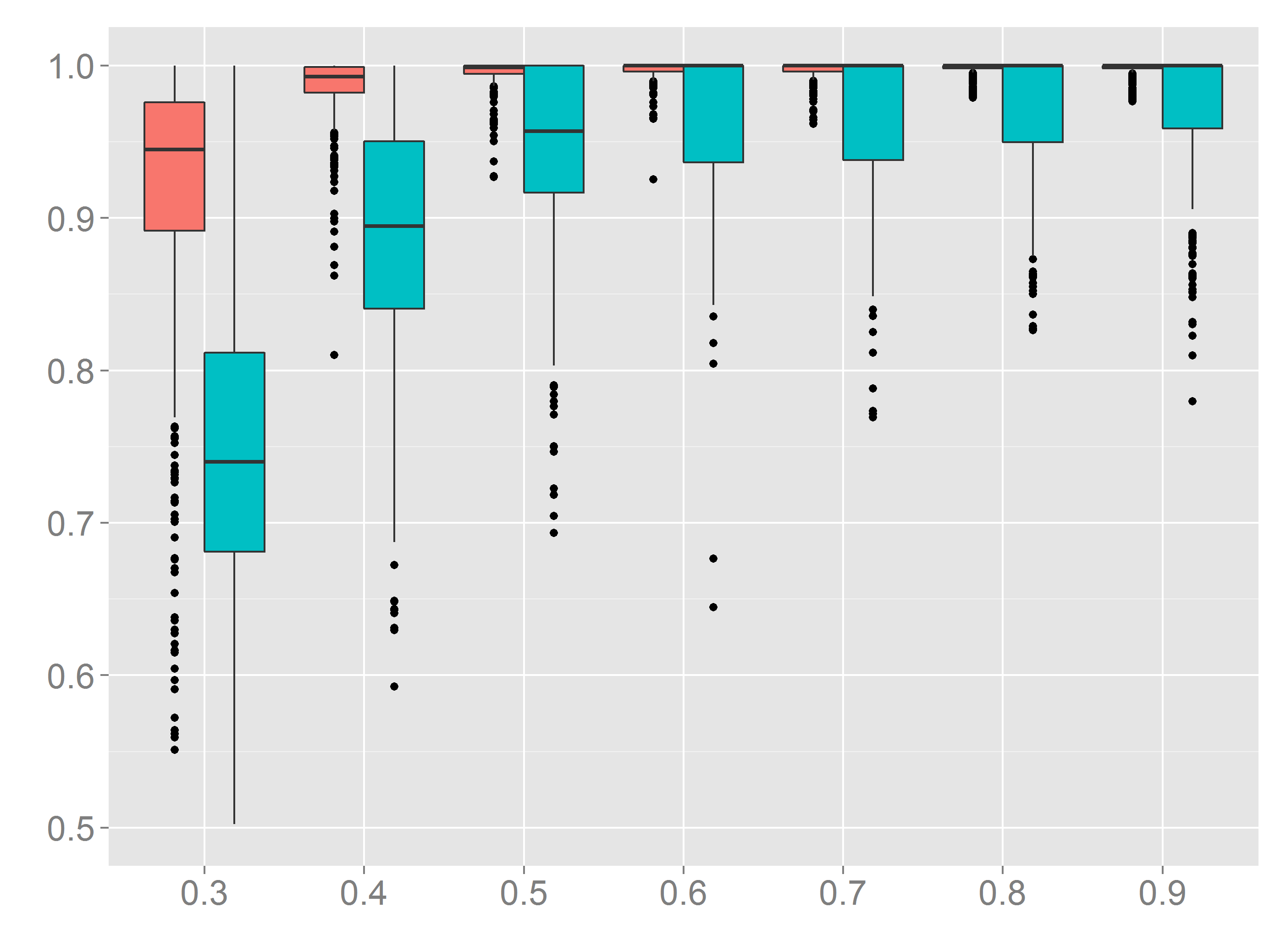}
	\end{center}
  \caption{AUC at the gene level (red) and region level (blue). The higher the AUC the better. Left: Simulation design 1 with fixed   $\rho_{1}=.5$ ($x$-axis: $\rho_0$). Right: Simulation design 2  ($x$-axis: $\rho_1$).}
  \label{fig:AUC}
\end{figure*}

Figure~\ref{fig:AUC} shows the AUC for the first simulation scenario
under various configurations, with $\rho_{1}$ fixed at $0.5$. When
$\rho_{0}$ is between 0.08 and 0.18, most regions are correctly detected. For $\rho_{0}=0.28$ (a value higher than what is observed on the reference dataset, see Figure~\ref{fig:rho0}), the task becomes difficult and the performances deteriorate.

On the second simulation study, the behavior of \SegCorr was explored under different $\rho_{1}$. Obviously the
task becomes easier when $\rho_{1}$ gets larger. Figure~\ref{fig:AUC}
shows that \SegCorr performs well when $0.5\leq \rho_{1} \leq
0.9$. When $\rho_{1} \leq 0.5$, (remind that the background
correlation can be as high as $0.2$, see Figure~\ref{fig:rho0})
although the performances remain good at the gene level, the
boundaries of the regions are detected less accurately.

\subsection{Comparison with the TCM algorithm}

\input{SimTCM.tex}

%% file: SimTCM.tex
\begin{figure*}
\begin{center}
    \includegraphics[width=.45\textwidth, height=0.5\textwidth]{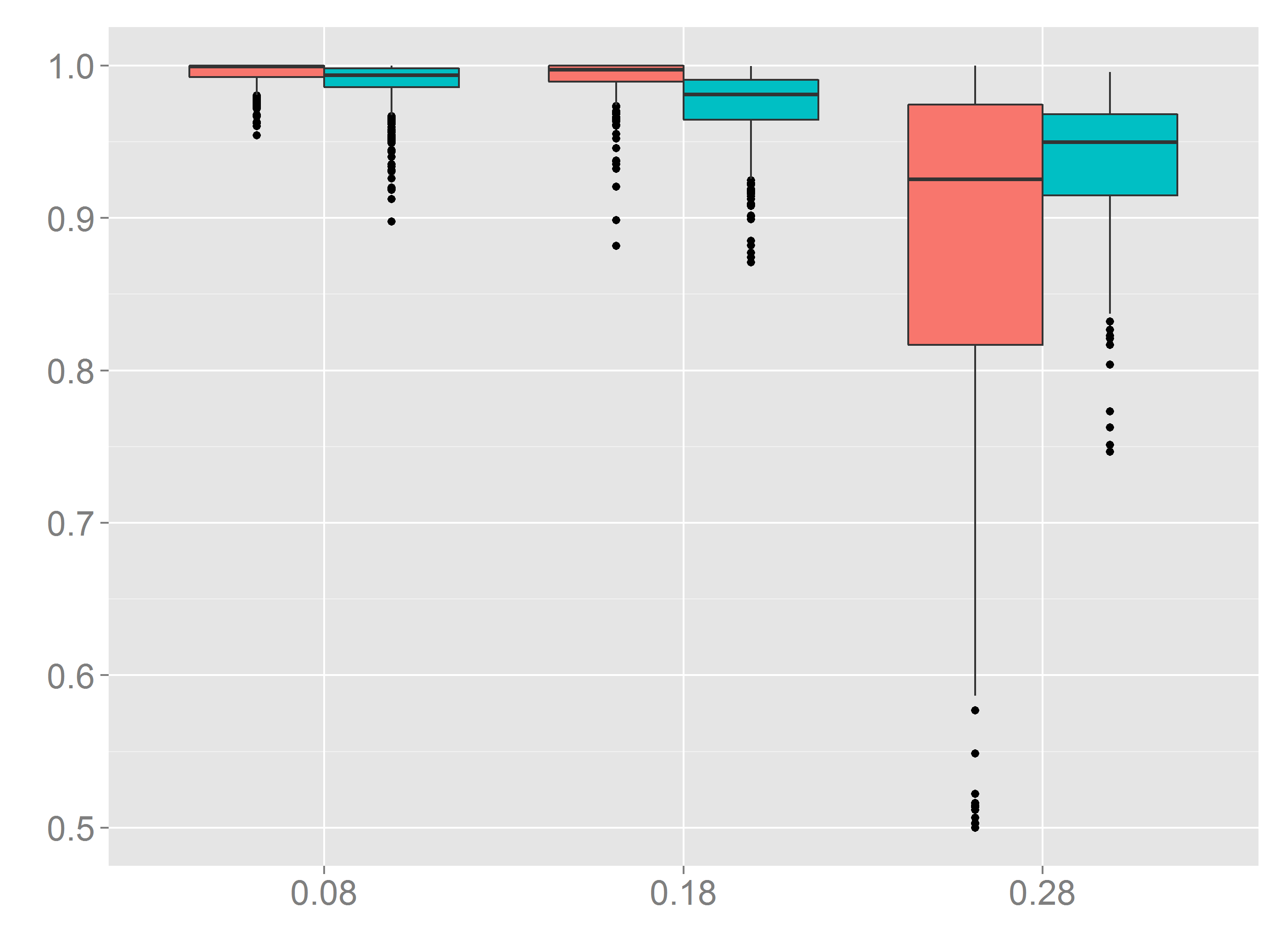}
    \includegraphics[width=.45\textwidth, height=0.5\textwidth]{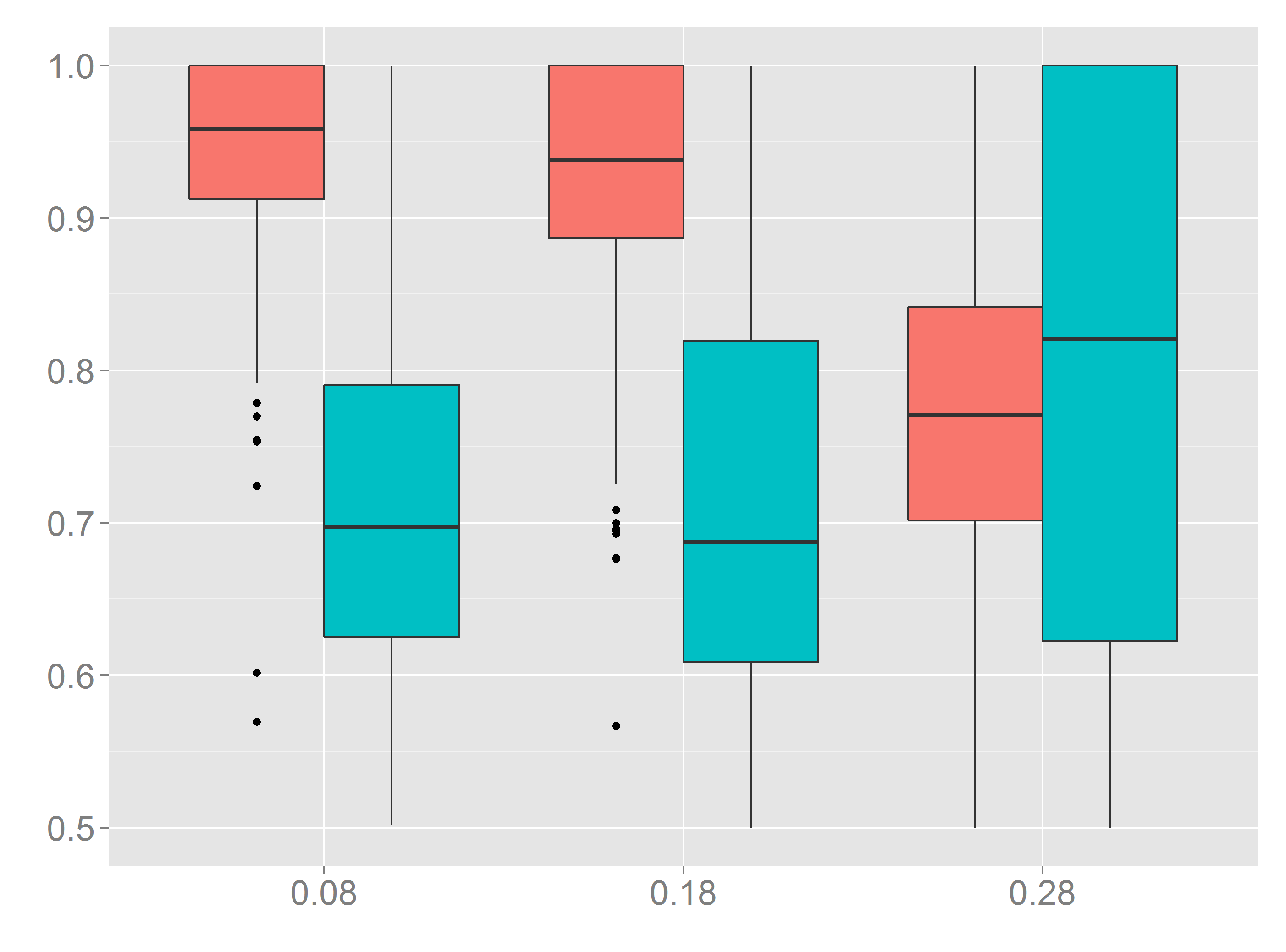}
\end{center}
  \caption{AUC of the \SegCorr (red) and TCM (blue) algorithms for the simulation scenario 1 as a function of $\rho_0$. Left: gene level. Right: region level.\label{fig:tcm}}
\end{figure*}

\SegCorr was compared with the TCM algorithm introduced
by \citet{reyal2005visualizing} for the detection of local
correlations. In the literature, many methods tackling the same problem as the TCM have been proposed.  The choice of the TCM as a competing method was based on the availability of the code. 
Figure~\ref{fig:tcm} displays the AUC achieved by \SegCorr and TCM
under Scenario 1. When $\rho_0$ is large ($\rho_0 = 0.28$), one observes that the mean
performance of both methods are comparable with higher variability
for \SegCorr at the gene level and at the region level for TCM. Since the aim is to detect regions rather that
genes, the \SegCorr procedure seems more appropriate. For small or medium values of background
correlations ($\rho_0 = 0.08, 0.18$) \SegCorr achieves
better AUC than TCM at both the gene and the region levels. As a conclusion, \SegCorr appears to be a
more consistent and efficient procedure to detect correlated regions.

%% file: DataPres.tex
\subsection{Data presentation} \label{sec:data_presentation}

{The dataset consists of $n=58$ bladder tumors.
Gene expression were measured using exon 1.0 Affymetrix arrays and RMA normalisation \citet{rma} was applied.
The number of genes per chromosome ranges from a 293 to 2192 (with average 950).
Additionally genomic and methylation data were collected for the same tumors.
Genomic data were obtained with Illumina Human610-Quad SNP arrays
and methylation data with Illumina Human methylation 450k  arrays.
For the latter, the normalization procedure proposed by \citet{teschendorff2013beta} was used. The combination of probe positions was made according to the custom BrainArray EntrezGene annotation \citep{dai2005evolving}.}

%% file: SRobustness.tex
\subsection{Study of the model selection threshold $S$}
\label{sec:SRobustness}

\begin{figure*}
\begin{center}
\includegraphics[width=0.45\textwidth, height=0.5\textwidth]{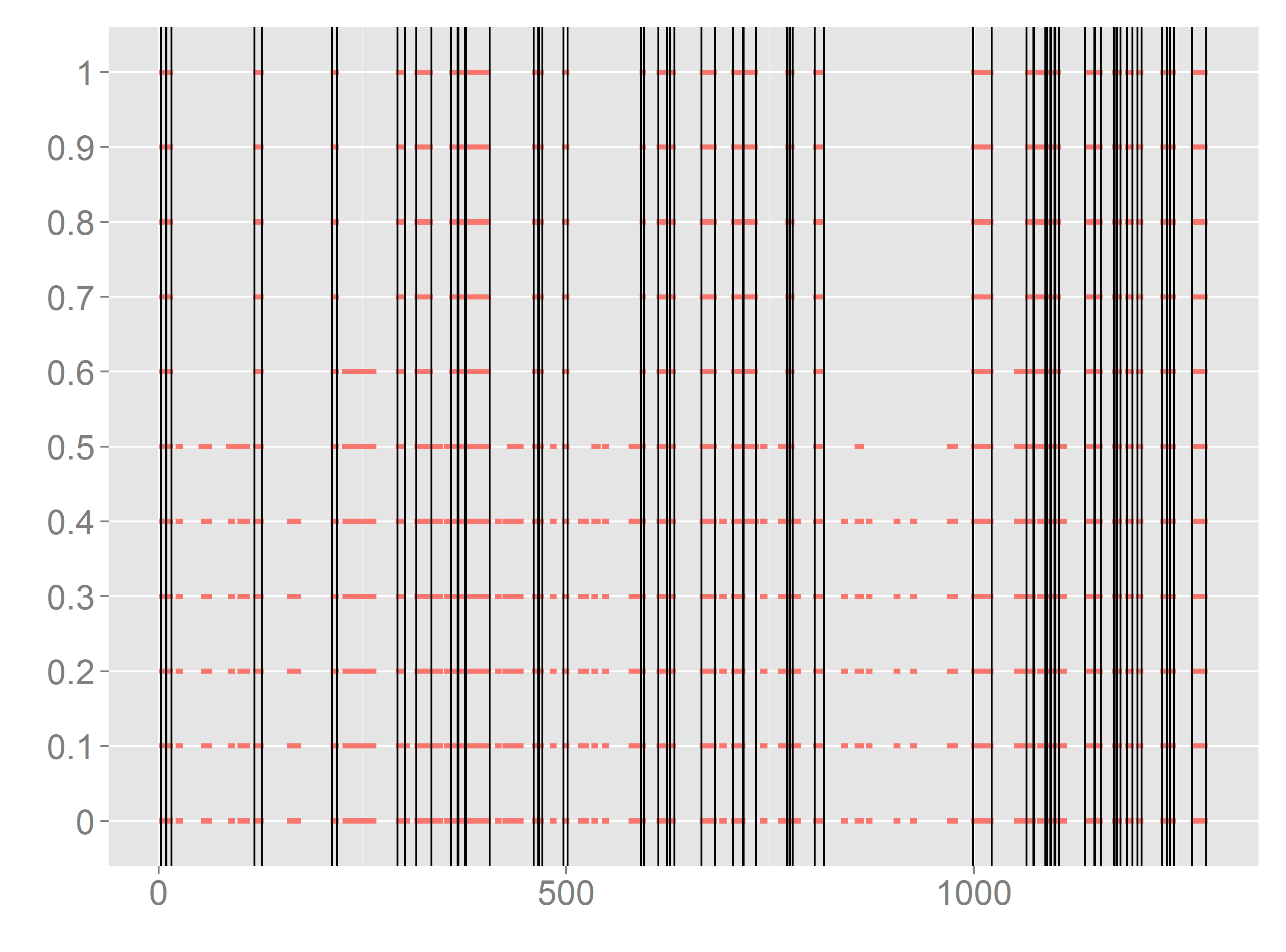}
\includegraphics[width=0.45\textwidth, height=0.5\textwidth]{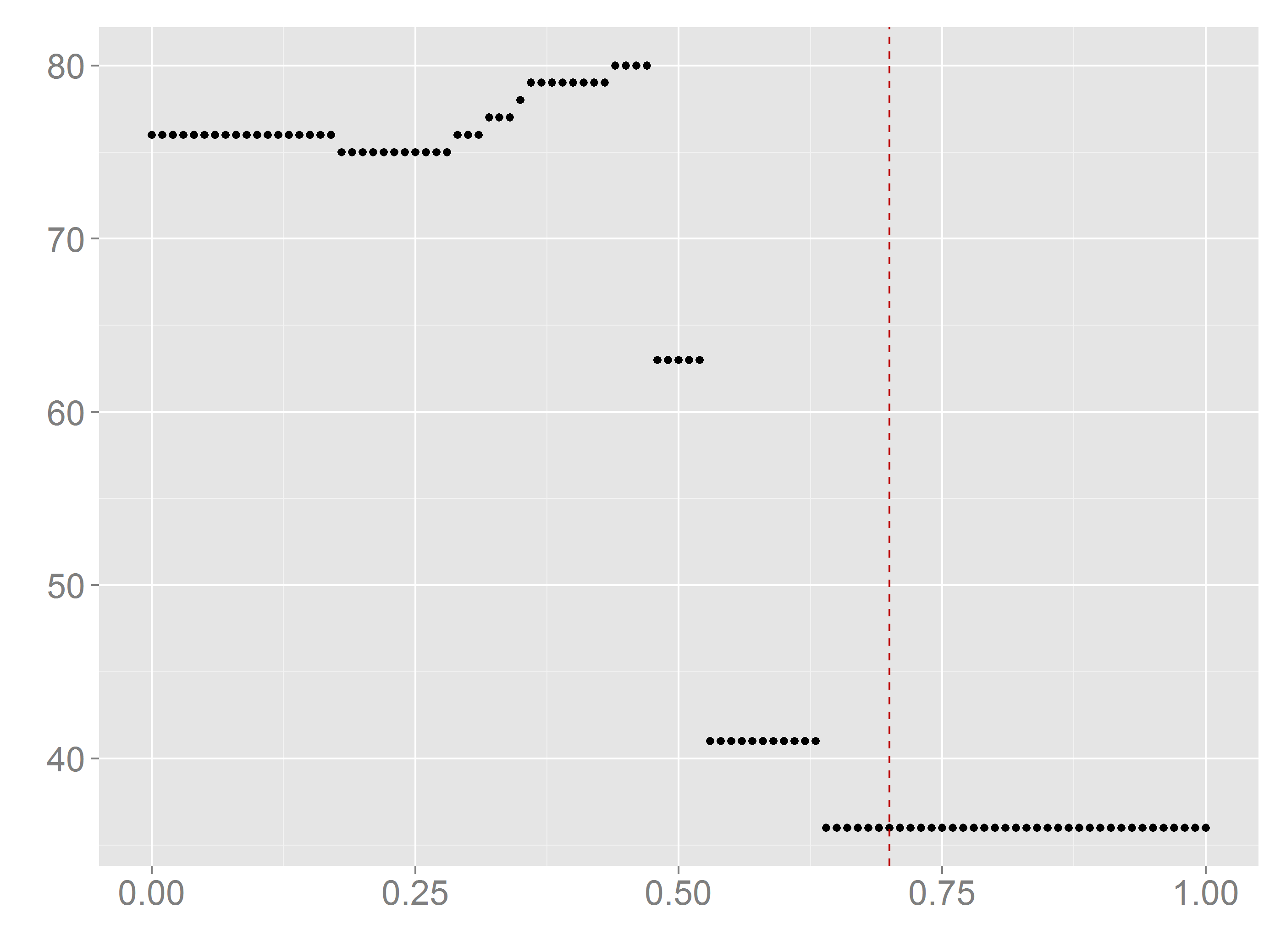}
\end{center}
\caption{Left: statistically significant regions in red obtained for
different values of $S$. The vertical lines correspond to the ones
obtained with the default value of $S$ we considered ($S=0.7$).
Right: number of statistically significant regions for different
values of $S$. The dotted vertical red line corresponds to $S=0.7$.}
\label{fig:chr3}
\end{figure*}

For the model selection criterion (see section
\ref{sec:regioninference}), the threshold~$S$ must be tuned in such
a way to avoid under/over-segmentation. The smaller the value of $S$ the higher the number of segments. As stated in Section \ref{modelselection},  $S$ was fixed to $0.7$ as advocated in \citet{Lavielle20051501}. Figure~\ref{fig:chr3} shows the evolution of the number and location of $H_1$ regions detected by SegCorr according to $S$ on a typical chromosome (chromosome 3). One can see that most of these $H_1$ regions are stable for values of $S$ between $0.6$ and $0.9$.

%% file: PrePro.tex
\subsection{Procedure for CNV correction} \label{sec:Preprocessing}
To correct the expression signal from CNV, one first needs to detect the CNV regions from the SNP signal. To this aim, we consider the segmentation method proposed by \citet{picard2011joint} implemented in the R package cghseg. Denote
$SNP_{it}$ the SNP signal of patient $i$ at position $t$, the model
writes
\begin{eqnarray} \label{sec:segmentationCNV}
SNP_{it}=\mu_{ik}+E_{it}\ \ \mbox{if $t \in
I_k^{i}=[t_{k-1}^{i}+1,t_{k}^{i}]$.}
\end{eqnarray}
where the $E_{it}$ are i.i.d centered Gaussian with variance
$\sigma^2$. The method estimates the number of regions, the
boundaries of the regions, denoted $\hat{t}_k^i$ and the signal
mean within each region $k$ in patient $i$, denoted
$\hat{\mu}_{ik}$.  \\
We then use the regression model \eqref{eq:regEXP1} to make the correction where $x_{ij}$ is the mean $\hat{\mu}_{ik}$ obtained
previously if the SNP position $\hat{t}_{k}^i$ corresponds to gene
$j$ of the expression signal in patient $i$. Since the SNP and expression signals are not aligned, there might be either one, many or no SNP probes that belong to the corresponding gene region. We then propose to define $x_{ij}$ as follows~: if one or many probes are related to gene $j$, mean $\hat{\mu}_{ik}$ or the
average of the different means is considered respectively; if there
is no probe, a linear interpolation is performed.

%% file: CNVregion.tex
\subsection{CNV Dependent Region} \label{study-correction-CNV}
\begin{figure}
  \includegraphics[width=0.45\textwidth, height=0.4\textwidth]{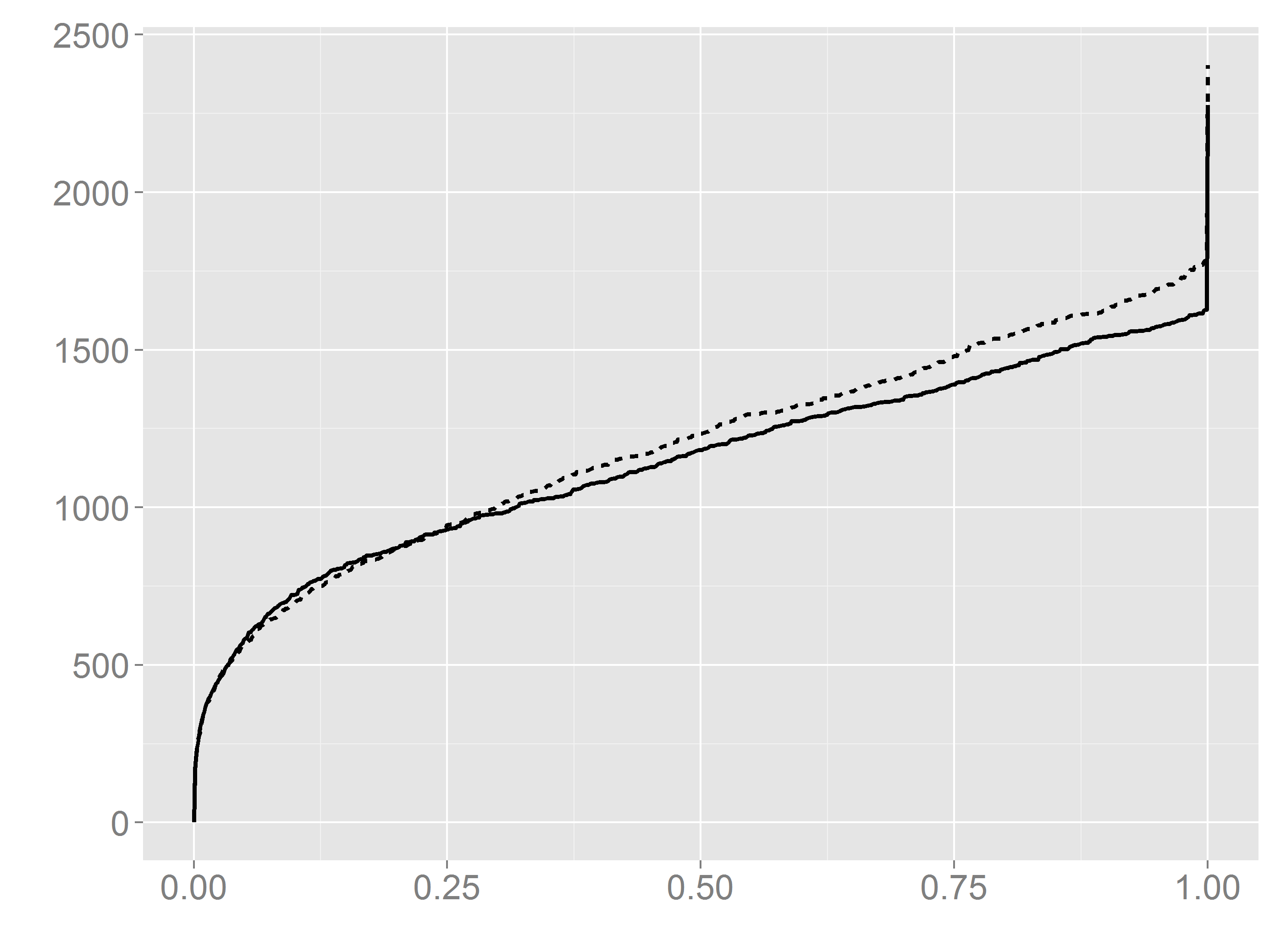}
  \includegraphics[width=0.45\textwidth, height=0.4\textwidth]{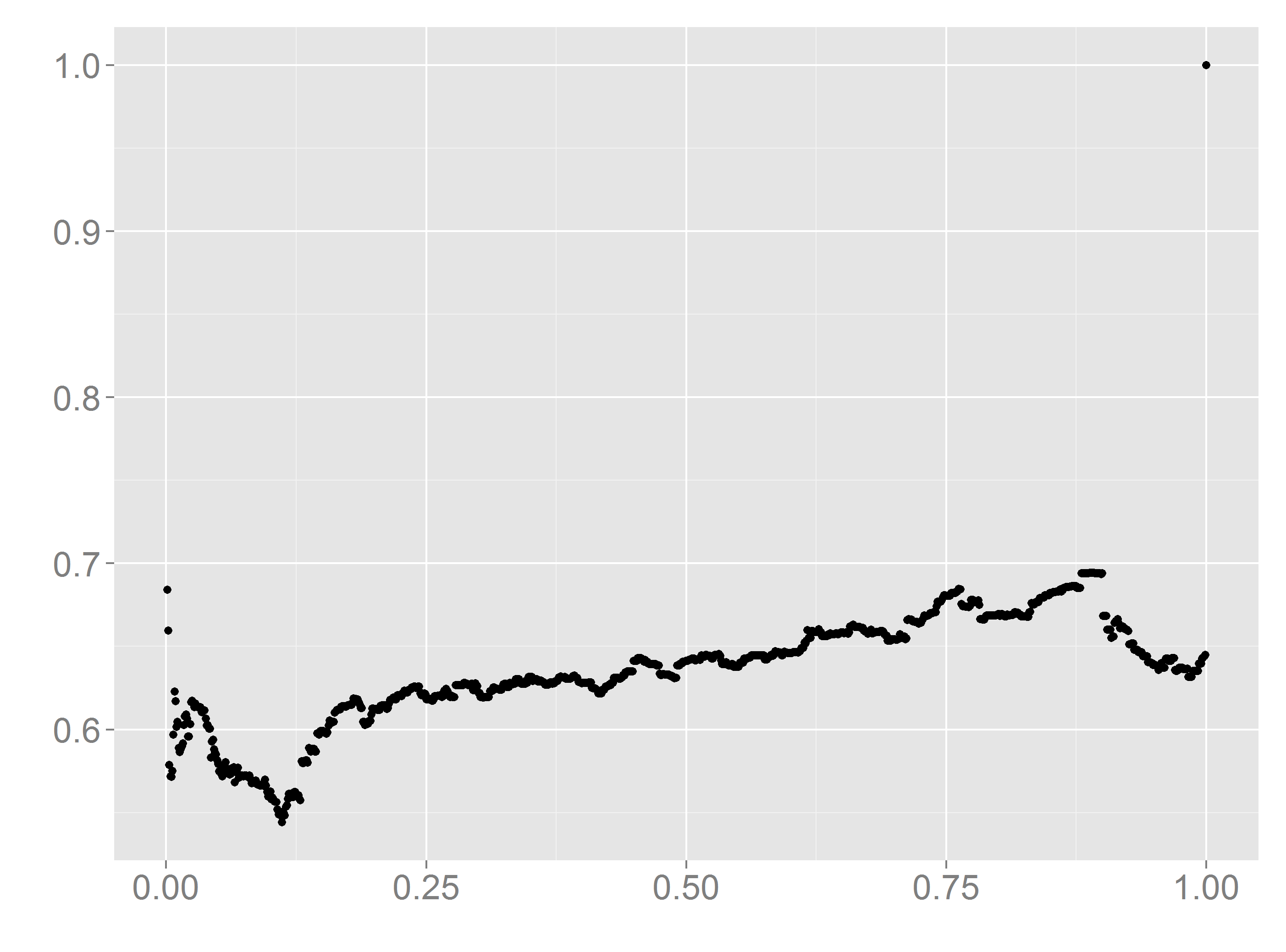}
  \caption{Left: Number of statistically significant regions as a function of $\alpha$ (solid line: corrected signal, dotted line: raw signal). Right: proportion of significant genes common in the two signal as a function of $\alpha$.}
  \label{fig:comparison}
\end{figure}

We first investigate the effect of CNV correction (described in Section \ref{sec:Preprocessing}) by comparing the results obtained on the raw and corrected signals.
Figure~\ref{fig:comparison} displays the number of significant $H_1$ regions as a function of the test level $\alpha$ for both the raw and corrected signals. For small values of $\alpha$ (which are typically used for testing significance), the number of detected regions are quite similar. However, only one third of the detected genes are common, meaning that the regions detected with the two signals are quite different. Furthermore, as the correction remove all effects due to CNV, the estimated background correlation is lower in the corrected signal than in the raw signal (mean decrease across all chromosomes of $.02$). This makes the test we propose more powerful and explains why, while CNV-due regions are removed, the number of detected regions for given $\alpha$ remains about the same. 

To illustrate this phenomenon more precisely, we considered a set of four regions in chromosomes 2, 6, 12 and 16 known to be associated with CNV in bladder cancer \citep{heidenblad,cancer2014comprehensive}. These regions, given in Table \ref{Tab}, are detected by \SegCorr when applied to the raw expression data. When considering the corrected signal, these regions are not detected any more. 
For example, when considering the region in chromosome 2, the background correlation was $\widehat{\rho}_0 = 0.163$ and the correlation within this region was $\widehat{\rho}_k = 0.561$,  resulting in a highly significant $p$-value: 2.1e-7. After correction we get $\widehat{\rho}_0 = 0.132$ and $\widehat{\rho}_k = 0.284$, which results in a non-significant $p$-value: 2.5e-2.

\begin{table}
  \begin{center}
   \begin{tabular}{cl}
 Chrom. & Genes \\
  \hline
  2 & \footnotesize{ASAP2, ITGB1BP1, CPSF3, IAH1, ADAM17, YWHAQ, TAF1B} \\
  6 & \footnotesize{MBOAT1, E2F3, CDKAL1, SOX4, LINC00340} \\
  12 & \footnotesize{MDM2, CPM, CPSF6, LYZ, YEATS4, FRS2, CCT2} \\
  16 & \footnotesize{COG7, GGA2, EARS2, UBFD1, NDUFAB1,  PALB2,  DCTN5} \\
  \end{tabular}
  \caption{Four examples of CNV-dependent regions \label{Tab}}
  \end{center}
\end{table}

More generally, over the 184 regions solely detected on the raw signal with p-value smaller than 5\% (before multiple testing correction), more than a half (98) get non significant when considering the corrected signal. This explains a substantial part of the difference between the regions detected on raw and corrected signals. This also shows that the proposed CNV correction strategy performs reasonably well.


{
}

%% file: MethylatedRegion.tex
\subsection{{CNV-independent regions}}\label{methylation}


\paragraph{General description.}
When applied to the CNV-corrected expression signal, \SegCorr detected 569 significant  regions ($p$-value adjusted $\leq 0.05$) which are distributed throughout the genome (an average of 23~regions per chromosome). Among these regions, 158~regions contained well known gene family clusters such as the HOXA, HOXB, HOXD clusters,  several KRT clusters, the epidermal differentiation complex,  and HLA gene families clusters whose expression is known to be co-regulated \citep{sproul2005role}. We next undertook a Gene Ontology terms analysis and identified an enrichment of genes belongs to the keratinization pathway ($p$-value 4.09E-19 and FDR $q$-value 9.01E-16). The expression of this pathways is strongly associated with a subgroup of bladder cancer called basal-like bladder cancer \citep{rebouissou2014egfr}. 11/28 CNV independent regions detected by TCM using different platforms (Affymetrix U95A for the transcriptome and BAC arrays for the genomic alterations) were also detected by \SegCorr \citep{stransky2006regional}.

\begin{figure*}
	\begin{center}
  \includegraphics[width=0.45\textwidth, height=0.4\textwidth]{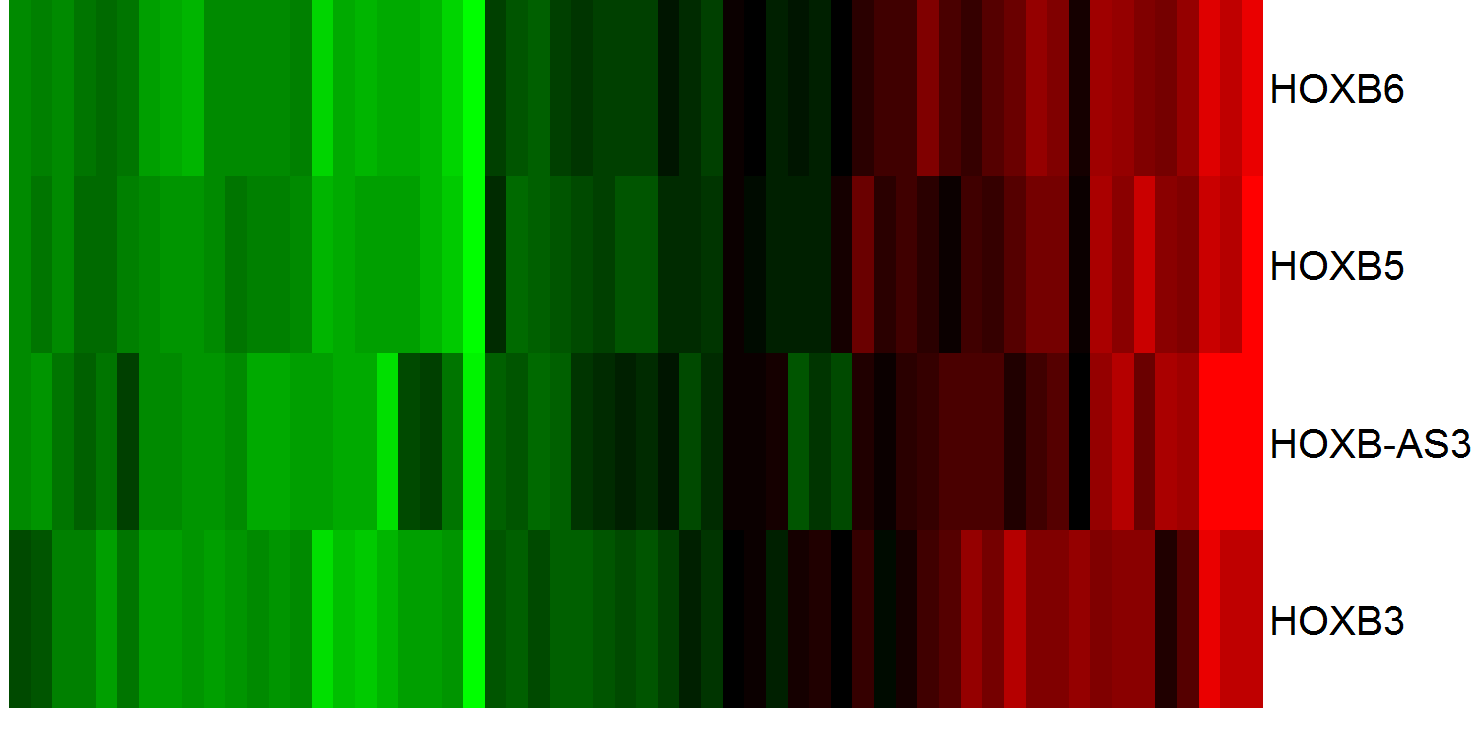}
  \includegraphics[width=0.45\textwidth, height=0.4\textwidth]{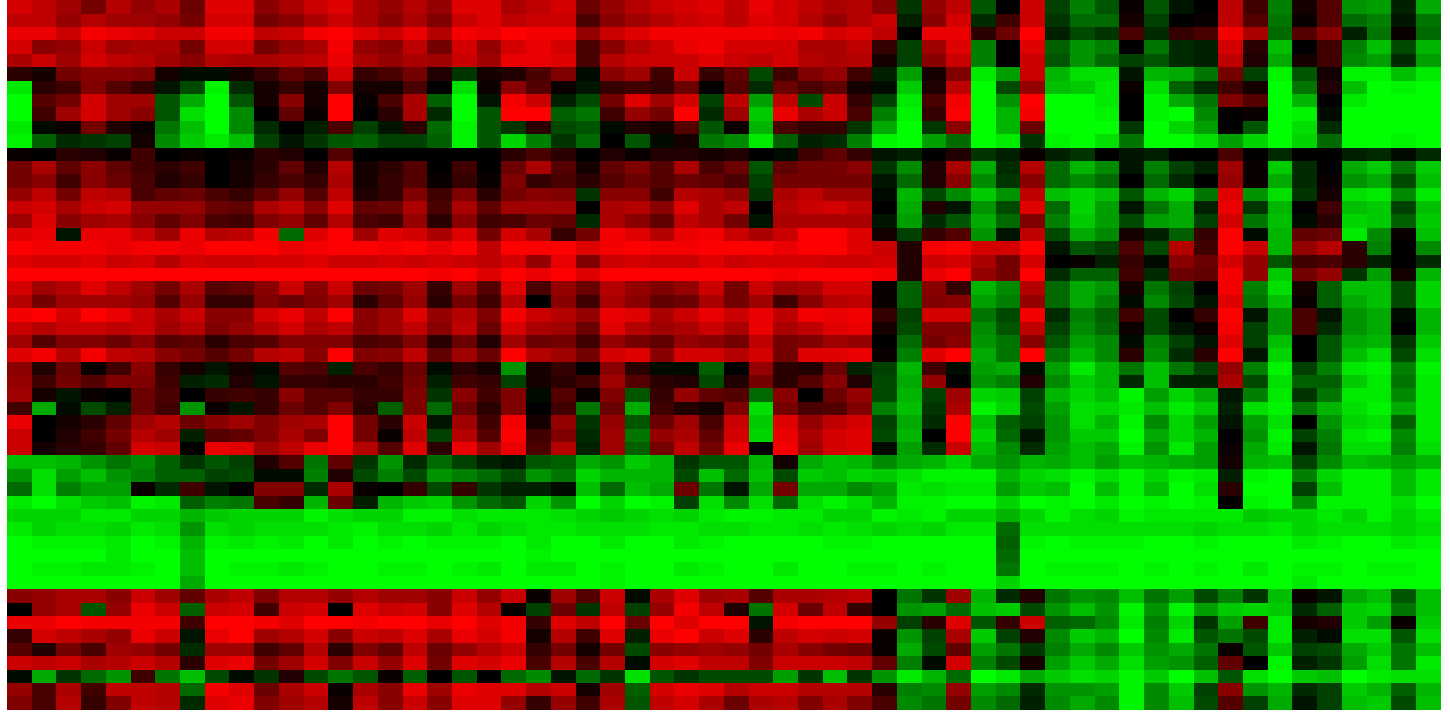}
	\end{center}
\caption{
  Expression (left) and methylation (right) data from Region 17-7. The ordering of the patients (x-axis) is kept the same in the two plots. }
  \label{fig:177GE}
\end{figure*}

\begin{figure}
\begin{center}
    \includegraphics[width=0.7\textwidth, height=0.5\textwidth]{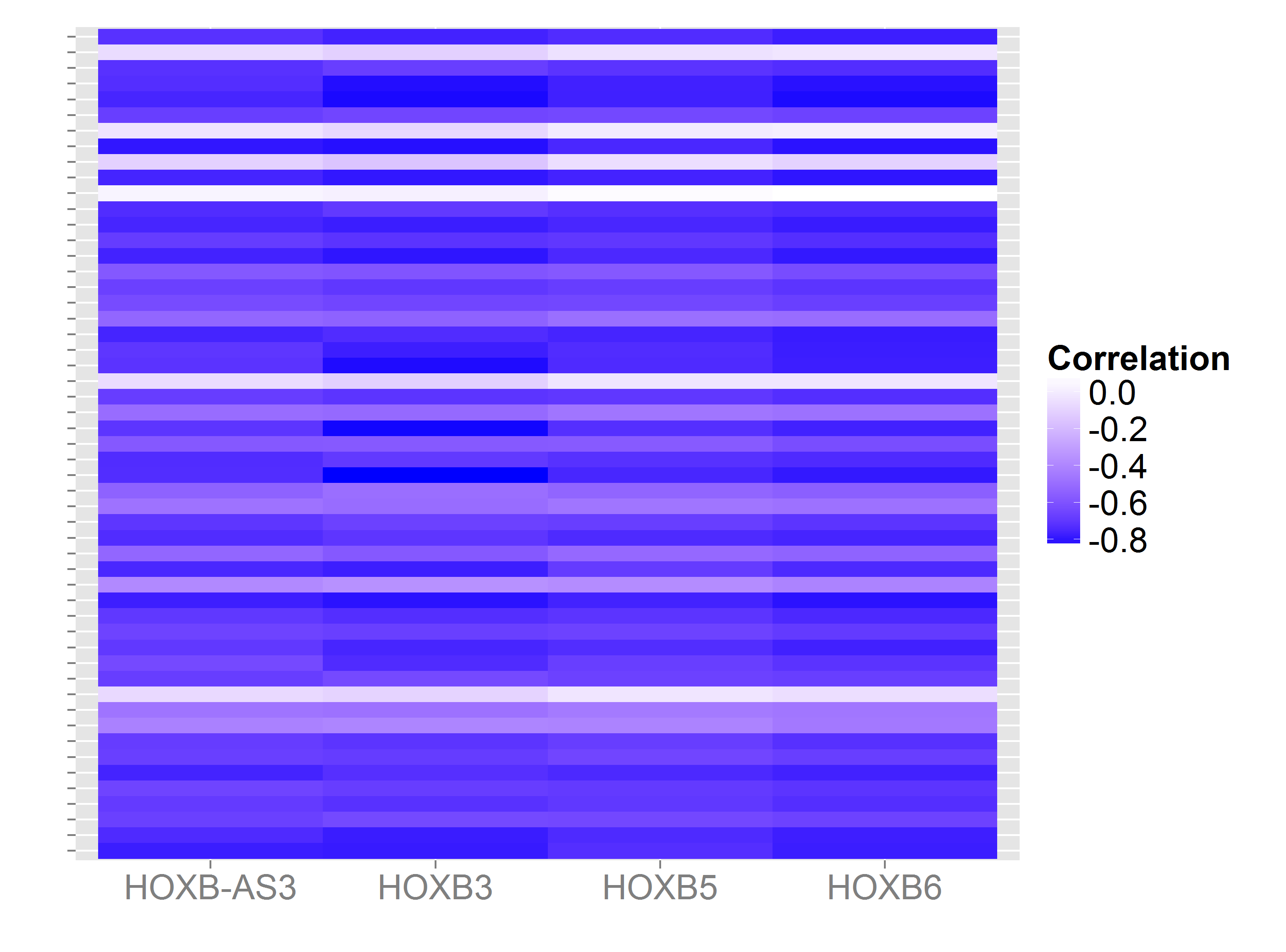}
		\end{center}
    \caption{
    Correlations between expression and methylation data from Region~$17$-$7$.} \label{fig:177MER}
\end{figure}

\paragraph{An example of epigenetic region.}
We now present a region where the observed correlation is not due to CNV but can be associated with an epigenetic mark. When applied to the CNV corrected expression data, \SegCorr detects a region of four genes (HOXB3, HOXB-AS3, HOXB5, HOXB6: $\widehat{\rho}_{k}= 0.93$, $p$-value = 3.1e-8) in chromosome 17. This region has already been studied by \citet{vallot2011novel} and has been referred to as 17-7.\\
Figure~\ref{fig:177GE} (left) shows a clear pattern in the expression data, which is detected by \SegCorr. The right panel provides the DNA methylation data for the same region, which also depicts a clear pattern. This suggest that this region is silenced by an epigenetic mechanism associated with DNA methylation. \\
This statement is supported by Figure~\ref{fig:177MER} which depicts the correlation between each gene expression and the methylation signal at each locus within the region. It shows a large proportion methylation sites being negatively correlated with the expression of the four genes. This indicates that high DNA methylation level in this region is 
associated with the silencing of these genes.

%% file: Discussion.tex
The identification of co-regulated chromosomal regions has many
implications in biology. In this paper, we developed a method to
identify these regions and we applied it to cancer data. The method
relies on a formal definition of what correlated regions are. It takes
advantage of an efficient algorithm and a statistical test, which are
both exact. We also proposed a correction strategy that allows us to
investigate the possible causes of the observed correlations. \\
Using this method, we could identify copy number dependent and copy
number independent correlated regions of expression. Copy number
dependent regions correspond to genomic alterations; copy number
independent regions could be due to different mechanisms, including
epigenetic mechanism. We showed, for one region, which is part of the
HOXB complex, that there is negative correlation between expression
and DNA methylation. The detected regions should be further
investigated to better understand the underlying mechanism. \\
While the expression data used here were acquired using the microarray technology, any other technology, including RNA-seq, can be used as well.



In our analysis, we have assumed stretches of correlated contiguous
neighboring genes. This is obviously a simplification. Within a
correlated region, a gene (or a few genes) could exhibit a weak or
even a negative correlation with the other genes. This could occur for
different reasons: the gene can be not expressed; alternatively, the gene could be non
affected by the regulation process that impacts the other ones;
finally, the gene could be impacted in a opposite way compared with
the other ones. Note that genes that exhibit no expression or no
variation in the dataset can be detected and could be discarded
before applying the analysis. While this preprocessing was not
required in the present study, running the analysis without removing
non-expressed genes would lower the performance of any method aimed at
finding correlated (and reasonably homogeneous)
regions. Alternatively, looking for the effect of adding a variable
number of uncorrelated genes in correlated regions is an obvious
follow-up of the present work.

The proposed correction strategy could easily be generalized to more
than one signal to correct for, as it does not rely on a joint
modeling of all types of data at hand. Furthermore the segmentation
used in the correction step enables us to deal with signals with
different probe densities. Finally, this correction approach allowed
us to keep all tumors in the study, as opposed
to \citep{stransky2006regional} were tumors with CNV in a given region were excluded when analysing this region. \\
Also, prior information on genes or regions could be accounted for in the segmentation step. Indeed, the likelihood $\widehat{\Lcal}(\tau, \tau')$ associated with a given region can be reweighted or penalized, the dynamic programming algorithm then applies with the same computational complexity.

%% file: Appendix.tex
\subsection{Proof of Lemma \eqref{lem:rho}}\label{annex: lemma}

Throughout this proof, we drop index $k$ for sake of clarity. For a region with length $\ell$, the covariance matrix $\Sigma$ in Equation~(\ref{eq:CorrelationModel}) can be rewritten as:
\begin{displaymath}
\Sigma = (1-\rho)I + \rho J
\end{displaymath}
where $I$ stand for the $\ell \times \ell$ identity matrix and $J$ for the $\ell \times \ell$ matrix with all entries equal to one.
The inverse of this matrix has the form $\Sigma^{-1} = aI + bJ$ where
$$
a =\frac{1}{1-\rho},
\qquad
b = - \frac{\rho}{(1-\rho)\left[1 -\rho +\ell\rho \right]}.
$$
The determinant of $\Sigma$ is
\begin{equation*} 
|\Sigma| = (1-\rho)^{\ell-1}(1 -\rho +\ell\rho )
\end{equation*}
and the trace term $\mbox{tr}\left[ \nECNV \Sigma^{-1} (\nECNV)^\top \right]$ in equation \eqref{eq:Lk} yields
$$
\mbox{tr}\left(\ECNV(aI+bJ)\TECNV\right) = an\ell  + bn \sum_{j=1}^{\ell}{\sum_{k=1}^{\ell}{\hat{G}_{jk}}}.
$$
Combining all the above gives the log-likelihood for this region:
\begin{eqnarray*}
-2\log{\Lcal} &=& n \left[\log{(1-\rho+\ell\rho)} + (\ell - 1)\log{(1-\rho)} \right] \\
&&+\frac{n\ell}{1-\rho} - \frac{\rho n\sum_{j}^{\ell}{\sum^{\ell}{\hat{G}_{jk}}}}{(1-\rho)\left[1 -\rho +\ell\rho \right]}.
\end{eqnarray*}
Optimizing this function wrt $\rho$ gives the formula of the MLE \eqref{eq:rhok_hat}. Pluging this estimate into the same function gives the contrast function given in  \eqref{eq:contrast}.

\subsection{Distribution of the test statistic}\label{annex: test}
Note $Y_i^{(k)}=(Y_{i,\tau_{k-1}+1},...,Y_{i,\tau_{k}})^T$. Using the same notations as in Section \ref{sec:regiontest}, one has
\begin{eqnarray*}
Y_i^{(k)} \sim \mathcal{N}\gp{0,\sigma_k} \Rightarrow Y_{i\bullet}^{(k)} \sim \mathcal{N}\gp{0,\frac{1+\gp{p_k-1}\rho_k}{p_k}}
\end{eqnarray*}
Because variables $Y_i^{(k)}$, $i=1,...,n$ are i.i.d so do variables $Y_{i\bullet}^{(k)}$, and consequently
\begin{eqnarray*}
\sum_{i}^{n}  \gp{Y^{(k)}_{i\bullet} - Y^{(k)}_{\bullet\bullet}}^{2} \sim \frac{1+\gp{p_k-1}\rho_k}{p_k} \chi^2_{n-1}
\end{eqnarray*}